\documentclass[twocolumn]{revtex4}

\usepackage{dcolumn}
\usepackage{bm}
\usepackage{amsmath}
\usepackage{graphicx}
\usepackage{amsmath,amssymb}
\usepackage{here}

\newcommand{\mean}[1]{\langle{#1}\rangle}

\newcommand{\ket}[1]{|{#1}\rangle}

\begin{document}

%\preprint{}

%%%%%%%%%%%%%%%%%%%%%%%%%%%%%%%%%%%%%%%%%%%%
%%%%%%%%%%%%%%%%%%%%%%%%%%%%%%%%%%%%%%%%%%%%
%%%%%%%%%%%%%%%%%%%%%%%%%%%%%%%%%%%%%%%%%%%%

\title{Coherent versus measurement feedback: \\
Linear systems theory for quantum information}



\author{Naoki Yamamoto}
\email[]{yamamoto@appi.keio.ac.jp}

%\thanks{}

\affiliation{
Department of Applied Physics and Physico-Informatics, 
Keio University, 
Hiyoshi 3-14-1, Kohoku, 
Yokohama 223-8522, 
Japan}


\date{\today}

%%%%%%%%%%%%%%%%%%%%%%%%%%%%%%%%%%%%%%%%%%%%
%%%%%%%%%%%%%%%%%%%%%%%%%%%%%%%%%%%%%%%%%%%%
%%%%%%%%%%%%%%%%%%%%%%%%%%%%%%%%%%%%%%%%%%%%

\begin{abstract} 

To control a quantum system via feedback, we generally have two options 
in choosing control scheme. 
One is the coherent feedback, which feeds the output field of the system, 
through a fully quantum device, back to manipulate the system without 
involving any measurement process. 
The other one is the measurement-based feedback, which measures the 
output field and performs a real-time manipulation on the system based 
on the measurement results. 
Both schemes have advantages/disadvantages, depending on the system 
and the control goal, hence their comparison in several situation 
is important. 
This paper considers a general open linear quantum system with the 
following specific control goals; 
back-action evasion (BAE), generation of a quantum non-demolished (QND) 
variable, and generation of a decoherence-free subsystem (DFS), all of 
which have important roles in quantum information science. 
Then some no-go theorems are proven, clarifying that those goals cannot 
be achieved by any measurement-based feedback control. 
On the other hand it is shown that, for each control goal, there exists a 
coherent feedback controller accomplishing the task. 
The key idea to obtain all the results is system theoretic characterizations 
of BAE, QND, and DFS in terms of controllability and observability 
properties or transfer functions of linear systems, which are consistent 
with their standard definitions. 

\end{abstract}



%\keywords{}


\maketitle


%%%%%%%%%%%%%%%%%%%%%%%%%%%%%%%%%%%%%%%%%%%%%
%%%%%%%%%%%%%%%%%%%%%%%%%%%%%%%%%%%%%%%%%%%%%
%%%%%%%%%%%%%%%%%%%%%%%%%%%%%%%%%%%%%%%%%%%%%

\section{Introduction}

{\it Should we perform measurement or not?} 
This question appears to be critical in quantum physics, particularly 
in quantum information science. 
For quantum computation, for instance, it is of essential importance to 
study differences between the conventional closed-system approach 
and the measurement-based one (i.e. the so-called one-way computation). 
This paper focuses on a specific aspect of this abstract and broad question; 
we will consider feedback control problems. 
That is, for a given open system (plant), we want to engineer another 
system (controller) connected to the plant so that the plant or the whole 
system behaves in a desirable way. 
The fundamental question is then, in our case, as follows; 
{\it should we measure the plant or not, for engineering a closed-loop 
system?} 
More precisely, in the former case, we measure the plant's output and 
engineer a classical controller that manipulates the plant using the 
measurement result -- this is called the 
{\it measurement-based feedback (MF)} approach. 
In the latter case, we do not measure it, but rather connect a 
fully quantum controller directly to the plant system in a feedback 
manner -- this is called the {\it coherent feedback (CF)} approach.

\begin{figure}[!h]
\centering 
\includegraphics[width=6.95cm]{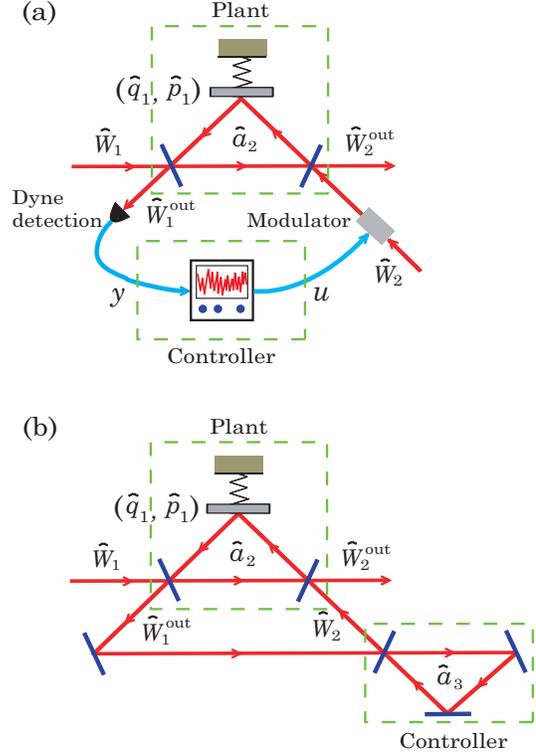}
\caption{
Example of (a) measurement-based feedback and (b) coherent feedback, 
for cooling a mechanical oscillator. 
}
\label{MF vs CF intro}
\end{figure}

A typical example is shown in Fig.~\ref{MF vs CF intro}; 
the plant is an open mechanical oscillator coupled to a ring-type optical cavity, 
and the control goal is to minimize the energy of the oscillator, or 
equivalently to cool the oscillator towards its motional ground state. 
As mentioned above, there are two feedback control strategies. 
One is the MF controller (Fig.~\ref{MF vs CF intro}~(a)) that measures the 
output field $\hat{W}^{\rm out}_1$ by for instance a homodyne detector; 
then, using the continuous-time measurement results $y(t)$, it produces 
the control signal $u(t)$ for modulating the input field $\hat{W}_2$. 
The other option is the CF control (Fig.~\ref{MF vs CF intro}~(b)), where 
we construct another fully quantum system that feeds the output field 
$\hat{W}^{\rm out}_1$ back to the input field $\hat{W}_2$, without 
involving any measurement component. 
The question is then about how to design a MF/CF controller that cools 
the oscillator most effectively.

Controller synthesis for a quantum system is in general non-trivial, but 
researchers' longstanding efforts have built a solid mathematical 
framework for dealing with those problems. 
For the MF case, actually there exists a beautiful {\it quantum feedback 
control theory} \cite{WisemanBook,BarchielliBook,Bouten2009} that was 
developed based on the 
{\it quantum filtering} \cite{Belavkin1992,Belavkin1993,Bouten2007} 
together with the classical control theory 
\cite{KailathBook,Wonham,Zhou Doyle book}. 
In fact, the above-described cooling problem can be formulated as a 
quantum {Linear Quadratic Gaussian (LQG)} feedback control problem 
and explicitly solved 
\cite{WisemanBook,Doherty1999,Hopkins,Mabuchi2013a,Mabuchi2013b}. 
Also the theory has been applied to various control problems in quantum 
information science such as error correction 
\cite{Ahn2002,MabuchiNJP2009,YamamotoPRA2012}. 
Notably, experiment of MF control is now within the reach of current 
technologies \cite{Haroche,Siddiqi,Devoret,Takahashi}. 
The CF control, on the other hand, has still a relatively young history 
though its initial concept was found in \cite{Wiseman1994 CF vs MF} 
back in 1994; 
but recently it has attracted increasing attention, leading as a result 
development of the basic control theory 
\cite{JamesTAC2008,NurdinSIAM2009,Gough2009TAC,GoughPRA2010} 
and applications 
\cite{Yanagisawa-2003,Gough2009,Kerckhoff-2010,Mabuchi-2011}. 
Some experimental demonstrations of CF control 
\cite{Mabuchi-2008,Iida,MabuchiOptExpress,KerckhoffPRX} also 
warrant special mention; 
in fact, one of the main advantages of CF is in its experimental 
feasibility compared to the MF approach.

Let us return to our question; which controller, MF or CF, is better? 
Now note that a CF controller is a fully quantum system whose random 
variables are in general represented by non-commutative operators, while 
a MF controller is a classical system with commutative random variables. 
Hence from a mathematical viewpoint the class of MF controllers 
is completely included in that of CF controllers. 
Thus our question is as follows; 
{\it in what situation is a CF controller better than a MF controller?} 
Actually there have been several studies exploring answers to this 
question \cite{Wiseman1994 CF vs MF,Nurdin-2009,Jacobs2012,
Mabuchi2013a,Mabuchi2013b}; 
most of these studies discussed problems of minimizing a certain cost 
function such as energy of an oscillator or the time required for 
state transfer. 
In particular in \cite{Mabuchi2013a,Mabuchi2013b}, the authors studied 
the problem discussed in the second paragraph and clarified that a certain 
CF controller outperforms any MF controller when the total mean phonon 
number of the oscillator is in the quantum regime; 
in other words, the two types of controllers do not show a clear difference 
in their performance for cooling, in a classical situation. 
This in more broad sense implies that a CF controller would outperform 
a MF controller only in a purely quantum regime. 
Consequently, our question can be regarded as a special case of the 
fundamental problem in physics asking in what situation a fully 
quantum device (such as a quantum computer) outperforms any 
classical one (such as a classical computer).

Towards shedding a new light on the above-mentioned fundamental 
problem, this paper attempts to clarify a boundary between the CF 
and MF controls for specific control problems. 
The problems are not what aim to minimize a cost function, but we will 
consider the following three; 
(i) realization of a back-action evasion (BAE) measurement, 
(ii) generation of a quantum non-demolished (QND) variable, 
and (iii) generation of a decoherence-free subsystem (DFS). 
The followings are brief descriptions of these notions in the input-output 
formalism \cite{GardinerBook,WallsMilburn}. 
First, if a measurement process is subjected only to a single noise quadrature 
(shot noise) and not to its conjugate (back-action noise), then it is called the 
BAE measurement \cite{BraginskyBook,Caves1980}; 
as a result BAE may beat the so-called standard quantum limit (SQL) and 
enables high-precision detection for a tiny signal such as a gravitational 
wave force. 
Next, a QND variable is a physical quantity that can be measured 
without being disturbed \cite{Braginsky1980}; 
more precisely, it is not affected by an input probe field but still appears 
in the output field, which can be thus measured repeatedly. 
Lastly, a DFS is a subsystem that is completely isolated from surrounding 
environment; 
that is, it is a subsystem whose variables are not affected by any input 
probe/environment field, and further, they do not appear in the corresponding 
output fields. 
Hence, a DFS can be used for quantum computation or memory 
\cite{ZanardiPRL1997,Lidar2003}. 
These three notions play crucial roles especially in quantum information 
science, thus their realizations are of essential importance. 
Indeed we find in the literature some feedback-based approaches 
realizing BAE \cite{Courty2003,Courty2003PRL,Vitali2004}, QND 
\cite{Wiseman1995}, and DFS \cite{Ticozzi2008,Ticozzi2009,
SchirmerPRA2010}.

\begin{figure}[t]
\centering 
\includegraphics[width=8.7cm]{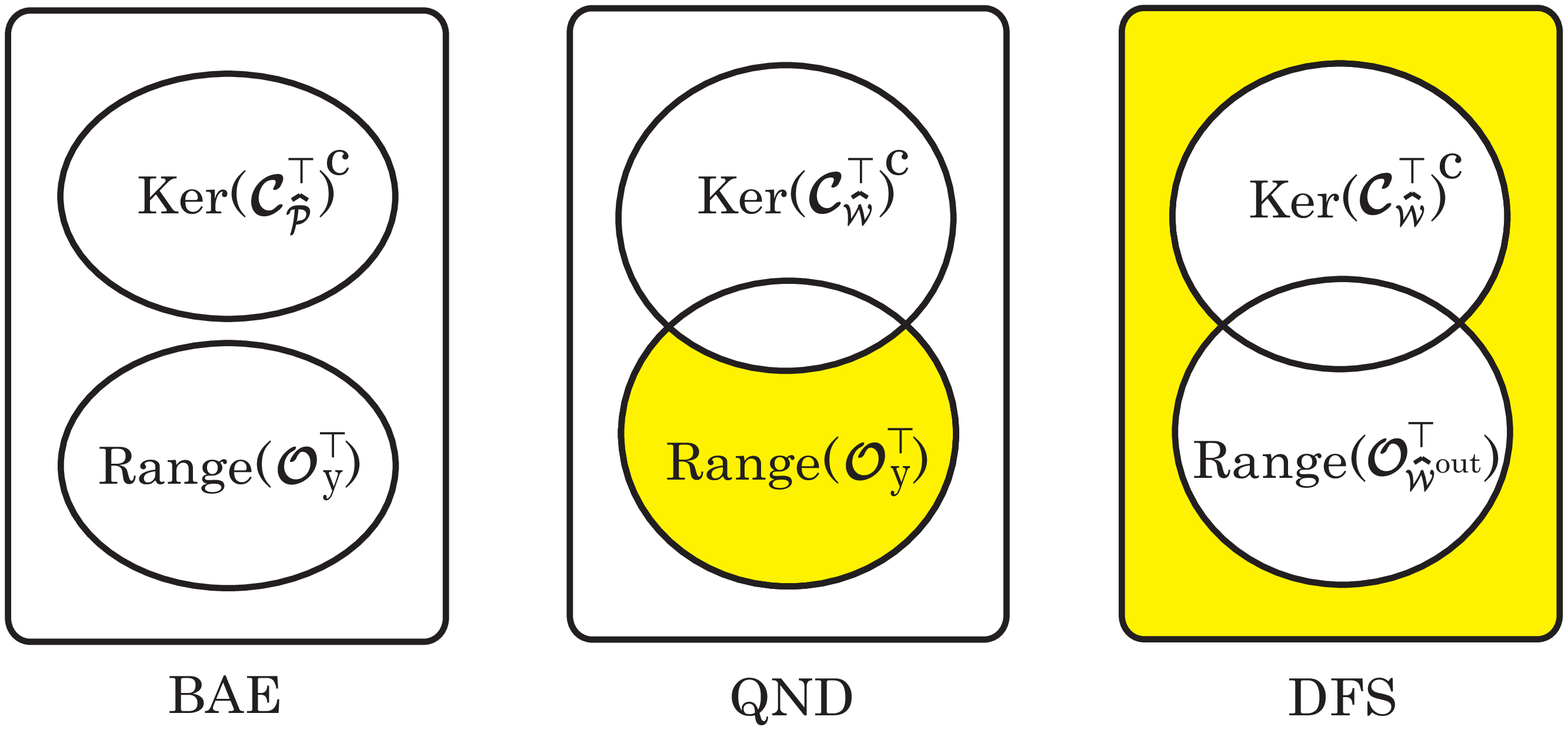}
\caption{
Systematic characterizations of BAE, QND, and DFS, represented by 
the set of vectors $v\in{\mathbb R}^{2n}$, where the corresponding 
quantum variables are given by $v^\top \hat x$. 
${\rm Ker}({\cal C}_\bullet^\top)^c$ and ${\rm Range}({\cal O}_\bullet^\top)$ 
denote the controllable and observable subspaces, respectively. 
The colored region represents the set of QND variables (middle) 
and the set of variables in a DFS (right). 
}
\label{Goals}
\end{figure}

Another feature of this paper is that we focus on general open {\it linear 
quantum systems} \cite{WisemanBook,GardinerBook,WallsMilburn}; 
this is a wide class of systems containing for instance optical devices 
\cite{BachorBook}, mechanical oscillators 
\cite{Mabuchi2013a,Mabuchi2013b,Courty2003,Courty2003PRL,
Vitali2004,LawPRA1995,Tsang2010,ClerkNJP2012,ClerkPRL2012,
WangScience,Chen2013,MiaoThesis}, and large atomic ensembles 
\cite{Duan2002,Kuzmich 2006,Parkins2006,Sorensen2007,PolzikRMP2010}. 
Linear systems are typical continuous-variables (CV) systems 
\cite{BraunsteinRMP,Furusawa2011}, which are applicable to several 
CV quantum information processing both in Gaussian case 
\cite{Ferraro2005,WeedbrookRMP2012} and non-Gaussian case 
\cite{Milburn2008,Khalili2010,YamamotoArxiv2014}. 
In both classical and quantum cases, for linear systems, the so-called 
{\it controllability} and {\it observability} properties can be well defined; 
further, those properties have equivalent representations in terms of a 
{\it transfer function}, which explicitly describes the relation between 
input and output. 
In fact a main advantage of focusing on linear systems is that we can 
have systematic characterizations of BAE, QND, and DFS in terms of 
the controllability and observability properties or transfer functions, 
which are consistent with the standard definitions found in the literature. 
Figure~\ref{Goals} is an at a glance overview of those characterizations, 
showing unification of the notions. 
Indeed this is the key idea to obtain all the results in this paper.

\begin{table}[t]
\caption{
The no-go theorems (left column). 
For both ``types" of control configurations, any MF controller cannot achieve 
the control goals; 
i.e. realization of BAE measurement, generation of a QND variable, 
and generation of a DFS. 
On the other hand, in every category we can find a CF controller 
achieving the task (right column). 
}
\centering 
\includegraphics[width=7.5cm]{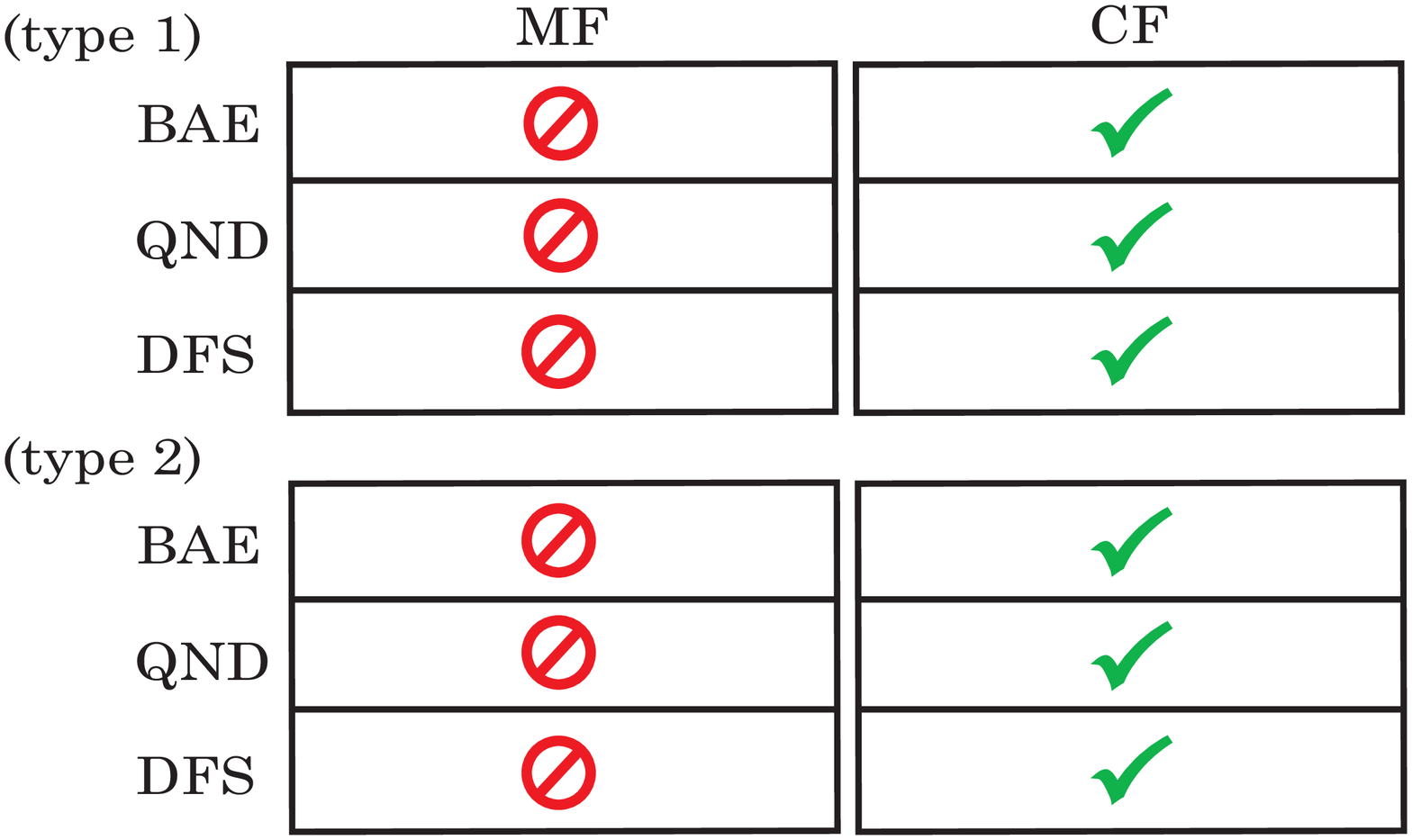}
\label{The no-go theorems}
\end{table}

Therefore our problem is, for a given open linear system, to design a CF/MF 
controller to realize BAE, QND, or DFS. 
For this problem, the results summarized in Table~\ref{The no-go theorems} 
are obtained. 
That is, no MF controller can achieve any of the control goals for general 
linear systems (there are two kinds of general configurations for feedback 
control, as indicated by ``type" in Table~\ref{The no-go theorems}). 
In contrast to these no-go theorems, for every category in the table we can find an example 
of CF controller achieving the goal. 
From the viewpoint of the above-mentioned fundamental question asking 
differences of the ability of quantum and classical devices, therefore, these 
results imply that BAE, QND, and DFS are the properties that can only be 
realized in a fully quantum device.

This paper is organized as follows. 
Section II reviews some useful facts in classical linear systems theory and 
describes a general linear quantum system with some examples. 
In Sec. III we discuss the three control goals, BAE, QND, and DFS, 
in the general input-output formalism and give their systematic 
characterizations in terms of the controllability-observability properties 
and also transfer functions; 
again, these new characterizations are special feature of this paper. 
Then the proofs of the no-go theorems are given in Secs. IV and V, each 
of which are devoted to the proofs for the type-1 and the type-2 MF 
control configuration, respectively. 
Sections VI and VII demonstrate systematic engineering of a CF 
controller achieving the control goal. 
In particular, in the type-2 case, we will study a Michelson's 
interferometer composed of two mechanical oscillators, which 
is used for gravitational wave detection.

{\bf Notations:} 
For a matrix $A$, the {\it kernel} and the {\it range} are defined by 
${\rm Ker}(A)=\{x \hspace{0.05cm}|\hspace{0.05cm} Ax=0\}$ and 
${\rm Range}(A)=\{y \hspace{0.05cm}|\hspace{0.05cm} y=Ax,~\forall x\}$, 
respectively. 
The complement of a linear space ${\cal X}$ is denoted by ${\cal X}^c$. 
$\emptyset$ means the null space. 
In this paper we do not use the terminology ``observable" to represent a 
measurable physical quantity (i.e. a self adjoint operator), because it has 
a different meaning in systems theory; 
a physical quantity is called a ``variable", e.g. a QND variable rather 
than a QND observable.

%%%%%%%%%%%%%%%%%%%%%%%%%%%%%%%%%%%%%%%%%%%%%
%%%%%%%%%%%%%%%%%%%%%%%%%%%%%%%%%%%%%%%%%%%%%
%%%%%%%%%%%%%%%%%%%%%%%%%%%%%%%%%%%%%%%%%%%%%

\section{Preliminaries: Linear systems theory and linear quantum systems}

%%%%%%%%%%%%%%%%%%%%%%%%%%%%%%%%%%%%%%%%%%%%%

\subsection{Linear systems theory}

A standard form of classical linear systems is given by 
\begin{equation}
\label{classical sys}
   \frac{dx}{dt}=Ax+Bu,~~~
   y=Cx. 
\end{equation}
$x(t)\in{\mathbb R}^n$ is a vector of $n$ c-number variables. 
$u(t)$ and $y(t)$ are vectors of real-valued input and output signals, 
respectively. 
$A, B$, and $C$ are real matrices with appropriate dimensions. 
In this paper, the following three questions are important; 
(i) which components of $x$ can be controlled by $u$, 
(ii) which components of $x$ can be observed from $y$, and 
(iii) in what condition $u$ does not appear in $y$? 
The answers are briefly described below. 
See \cite{KailathBook,Wonham,Zhou Doyle book} for more detailed 
discussion.

The first problem can be explicitly solved by examining the 
following {\it controllability matrix}: 
\begin{equation}
\label{classical cont matrix}
   {\cal C}_u = [B,~AB,~A^2B,\ldots,~A^{n-1}B]. 
\end{equation}
Indeed this matrix fully characterizes the controllable and uncontrollable 
variables with respect to (w.r.t.) $u(t)$. 
To see this fact, suppose $m=\dim{\rm Range}({\cal C}_u)<n$ and 
let $\{ d_i^{(1)}\}$ and $\{d_i^{(2)}\}$ be independent vectors spanning 
${\rm Range}({\cal C}_u)$ and ${\rm Range}({\cal C}_u)^c$, 
respectively. 
Further let us define $T_1=[d_1^{(1)},\ldots,d_{m}^{(1)}]$ and 
$T_2=[d_1^{(2)},\ldots,d_{n-m}^{(2)}]$. 
Then, as $A{\cal C}_u$ is spanned by $\{ d_i^{(1)}\}$, there exists 
a matrix $A_{11}$ satisfying $AT_1=T_1 A_{11}$. 
On the other hand $AT_2$ is in general spanned by all the vectors; i.e. 
$AT_2=T_1 A_{12}+T_2 A_{22}$. 
Note also that there exists a matrix $B_1$ satisfying $B=T_1B_1$. 
These relations are summarized in terms of the invertible square 
matrix $T=[T_1, T_2]$ as
\[
    AT=T \left[\begin{array}{cc}
                 A_{11} & A_{12} \\
                 0 & A_{22}
              \end{array}\right],~~
    B = T \left[\begin{array}{c}
                  B_1  \\
                  0 
              \end{array}\right]. 
\]
Thus the dynamics of $x'=T^{-1} x$ is given by  
\begin{equation}
\label{classical transformed sys}
   \frac{dx'}{dt} 
    = \left[\begin{array}{cc}
               A_{11} & A_{12} \\
                 0 & A_{22}
        \end{array}\right] x'
     + \left[\begin{array}{c}
               B_1 \\
               0 
            \end{array}\right] u. 
\end{equation}
Clearly $x'_2=V_2^\top x$ is free from $u$, where $[V_1, V_2]^\top=T^{-1}$; 
in particular, due to $V_2^\top T_1=0$, the uncontrollable variable $x'_2$ is 
characterized by ${\rm Range}(V_2) = {\rm Ker}({\cal C}_u^\top)$. 
Also the controllable one $x'_1=V_1^\top x$ is defined in 
${\rm Range}(V_1)={\rm Ker}({\cal C}_u^\top)^c$. 
Hence we call these sets the {\it uncontrollable subspace} and the 
{\it controllable subspace}, respectively
\footnote{
Usually the controllable and uncontrollable subspaces are defined by 
${\rm Range}({\cal C}_u)$ and ${\rm Range}({\cal C}_u)^c$, respectively. 
But in the quantum case a variable of interest is an infinite-dimensional 
operator and does not live in either of these subspaces; 
rather it is always of the form $v^\top \hat x$ and thus can be well 
characterized by the dual vector $v\in{\mathbb R}^{2n}$. 
This is the reason why we define the controllable and uncontrollable 
subspaces in the dual space as ${\rm Ker}({\cal C}_u^\top)^c$ and 
${\rm Ker}({\cal C}_u^\top)$, respectively. 
}. 
The following fact is especially useful in this paper: 
the system has an uncontrollable variable $r=v^\top x$ iff 
\begin{equation}
\label{classical QND}
    v \in {\rm Ker}({\cal C}_u^\top)
    ~~\Leftrightarrow~~
    v^\top A^k B=0,~~\forall k\geq 0. 
\end{equation}

The answer to the second question is obtained in a similar fashion. 
Let us define the {\it observability matrix}
\begin{equation}
\label{classical obser matrix}
   {\cal O}_y=[C^\top,~A^\top C^\top,~(A^2)^\top C^\top,\ldots,
      ~(A^{n-1})^\top C^\top]^\top. 
\end{equation}
Assume ${\rm dim}{\rm Ker}({\cal O}_y)=\ell<n$. 
Then, there exists a linear transformation 
$x\rightarrow x'=[x_1'\mbox{}^\top, x_2'\mbox{}^\top]^\top$ with 
$x'_2\in{\mathbb R}^\ell$ 
such that the system equations are of the following form: 
\begin{equation}
\label{classical transformed sys 2}
   \frac{dx'}{dt} 
     = \left[\begin{array}{cc}
               A_{11} & 0 \\
               A_{21} & A_{22}
            \end{array}\right] x'
     + \left[\begin{array}{c}
               B_1 \\
               B_2 
            \end{array}\right] u, ~~
     y = [C_1,~0] x'. 
\end{equation}
Thus $x'_1$ and $x'_2$ constitute the observable and unobservable 
subsystems w.r.t. $y$, respectively. 
The variables are represented by 
$x'_1=U_1^\top x$ with ${\rm Range}(U_1)
={\rm Range}({\cal O}_y^\top)$ and $x'_2=U_2^\top x$ with 
${\rm Range}(U_2)={\rm Range}({\cal O}_y^\top)^c$; 
as in the above case, we call these subspaces the {\it observable subspace} 
and {\it unobservable subspace}, respectively. 
In particular, there always exists a coordinate transformation such that 
$r=v^\top x$ is unobservable if and only if 
\begin{equation}
\label{classical QND 2}
    v \in {\rm Ker}({\cal O}_y)
    ~~\Leftrightarrow~~
    C A^k v =0,~~\forall k\geq 0. 
\end{equation}

The above two facts readily leads to the answer to the third question; 
that is, there is no subsystem that is controllable w.r.t. $u$ and 
observable w.r.t. $y$, which is algebraically represented by 
\begin{equation}
\label{classical BAE}
     {\rm Ker}({\cal C}_u^\top)^c \cap {\rm Range}({\cal O}_y^\top)=\emptyset
     ~~\Leftrightarrow~~
    CA^k B=0,~~\forall k\geq 0. 
\end{equation}
Note that this is further equivalent to 
${\rm Range}({\cal C}_u)\subseteq{\rm Ker}({\cal O}_y)$, which particularly 
implies $CT_1=0$ with $T_1$ defined below Eq.~\eqref{classical cont matrix}. 
Hence we have 
\[
   y=Cx=CTT^{-1} x=C[T_1,~T_2]x'=[0,~CT_2]x', 
\]
where $x'=T^{-1}x$. 
Together with Eq.~\eqref{classical transformed sys}, we now see that $u$ 
acts only on $x'_1=V_1^\top x$ while $x'_1$ is not visible from $y$; 
accordingly, $u$ does not appear in $y$.

The above conditions \eqref{classical QND}, \eqref{classical QND 2}, and 
\eqref{classical BAE} can be represented in terms of a {\it transfer function}; 
let us define the Laplace transformation of a time-varying signal $z(t)$ by 
\[
    z[s] := \int_0^\infty z(t)e^{-st}dt,~~~{\rm Re}(s)>0.
\]
In the Laplace domain, Eq.~\eqref{classical sys} is represented by 
$sx[s]=Ax[s]+Bu[s]$ and $y[s]=Cx[s]$, which consequently yield 
\[
    y[s]=\Xi_{u\rightarrow y}[s]u[s],~~~
    \Xi_{u\rightarrow y}[s]=C(sI-A)^{-1}B. 
\]
Thus,  the signal flow from $u$ to $y$ is explicitly characterized by 
the transfer function $\Xi_{u\rightarrow y}[s]$. 
We then readily see from the polynomial expansion of $\Xi_{u\rightarrow y}[s]$ 
w.r.t. $s$ that the condition \eqref{classical BAE} is equivalent to 
\begin{equation}
\label{classical BAE transfer fn}
       \Xi_{u\rightarrow y}[s] = 0,~\forall s.
\end{equation}
Likewise, Eqs.~\eqref{classical QND} and \eqref{classical QND 2} are 
respectively equivalent to
\begin{equation}
\label{classical QND transfer fn}
        \Xi_{u\rightarrow x'_2}[s] = 0,~\forall s~~~\mbox{and}~~~
        \Xi_{x'_2\rightarrow y}[s] = 0,~\forall s.
\end{equation}
%

%%%%%%%%%%%%%%%%%%%%%%%%%%%%%%%%%%%%%%%%%%%%

\subsection{Linear quantum systems}

In this paper, we consider a general open system composed of $n$ 
oscillators with canonical conjugate pairs $\hat q_i$ and $\hat p_i$ 
$(i=1, \ldots, n)$. 
Let us collect them into a single vector as 
$\hat x = [\hat q_1, \hat p_1, \ldots, \hat q_n, \hat p_n]^\top$. 
Then, the CCR $\hat q_i\hat p_j-\hat p_j\hat q_i=i \delta _{ij}$ (we assume 
$\hbar=1$) is represented by 
\begin{eqnarray}
\label{CCR}
& & \hspace*{-1.8em}
     \hat x \hat x ^\top -(\hat x \hat x^\top )^\top 
      = i \Sigma_n,
\nonumber \\ & & \hspace*{-1.7em}
     \Sigma_n = {\rm diag}\{\sigma, \ldots, \sigma \},~~
      \sigma = \left[\begin{array}{cc}
               0 & 1 \\
               -1 & 0
            \end{array}\right]. 
\end{eqnarray}
$\Sigma_n$ is a $2n\times 2n$ block diagonal matrix; 
we often omit the subscript $n$. 
The system is driven by the Hamiltonian
\[
     \hat H = \hat x^\top G \hat x/2,
\]
where $G=G^\top \in {\mathbb R}^{2n \times 2n}$. 
Further, it couples to environment/probe fields through the Hamiltonian 
$\hat H_{\rm int}=i \sum_j (\hat L_j \hat A_j^* - \hat L_j^* \hat A_j)$, 
where $\hat L_j = c_j^\top \hat x$ ($c_j \in {\mathbb C}^{2n}$, 
$j=1,\ldots,m$). 
Also $\hat A_j$ is the annihilation operator on the $j$th field, which 
under the Markovian approximation satisfies 
$[\hat A_i(t), \hat A_j^*(t')]=\delta_{ij}\delta(t-t')$; 
i.e. it is the white noise operator. 
Then, the Heisenberg equations of $\hat q_j$ and $\hat p_j$ are 
summarized to the following linear equation \cite{WisemanBook,
GardinerBook,WallsMilburn}: 
\begin{equation}
\label{linear dynamics}
   \frac{d\hat x}{dt} = A\hat x 
       + \Sigma_n C^\top \Sigma_m \hat {\cal W}. 
\end{equation}
The coefficient matrices are given by $A=\Sigma_n(G+C^\top\Sigma_m C/2)
\in {\mathbb R}^{2n\times 2n}$ (the second term is the Ito-correction term) 
and 
\[
    C = \sqrt{2}\big[ \Re(c_1), \Im(c_1), \ldots, \Re(c_m), \Im(c_m) \big]^\top
       \in {\mathbb R}^{2m\times 2n}. 
\]
Also we have defined 
$\hat {\cal W}=[\hat Q_1, \hat P_1, \dots, \hat Q_m, \hat P_m]^\top$, 
where 
\begin{equation}
\label{noise quadrature}
   \hat Q_j=(\hat A_j + \hat A_j^*)/\sqrt{2},~~~
   \hat P_j=(\hat A_j - \hat A_j^*)/\sqrt{2}i. 
\end{equation}
Further, the field variables change to 
\begin{equation}
\label{linear output}
   \hat {\cal W}^{\rm out} 
     = C\hat x + \hat {\cal W}.
\end{equation}
The set of equations \eqref{linear dynamics} and \eqref{linear output} is 
the most general form of open {\it linear quantum systems}.

All the $2m$ elements of the vector $\hat{\cal W}^{\rm out}$ in 
Eq.~\eqref{linear output} cannot be measured simultaneously, 
because they do not commute with each other. 
In fact, without introducing additional noise fields as explained just later, 
we can measure only at most half of them; 
that is, the output equation associated with a {\it linear measurement}, 
which is realized by a Homodyne detector, is of the form 
\begin{equation}
\label{measurement}
   y = M_1 \hat{\cal W}^{\rm out} 
        = M_1 C \hat x + M_1 \hat{\cal W},
\end{equation}
where $M_1$ is a $m\times 2m$ real matrix satisfying 
$M_1 \Sigma_m M_1^\top=0$ and $M_1 M_1^\top=I$. 
Actually, all the elements of $y(t)$ are classical signals commuting 
with each other as well as with those of $y(t')$ for all times $t, t'$; i.e. 
\[
    [y_i(t), y_j(t')] = 0,~\forall i, j, ~\forall t, t'. 
\]
Let us further introduce $\bar{y}=M_2 \hat{\cal W}^{\rm out}$ with 
matrix $M_2$ such that $M^\top=[M_1^\top, M_2^\top]$ is a symplectic 
and orthogonal matrix, which as a result leads to 
\begin{eqnarray}
& & \hspace*{0em}
\label{M1 M2 conditions}
    M_2 \Sigma_m M_2^\top=0,~~
    M_2 M_2^\top=I,~~
    M_1 \Sigma_m M_2^\top=I, 
\nonumber \\ & & \hspace*{0em}
    M_1M_2^\top=0,~~
    M_1^\top M_1 + M_2^\top M_2 = I. 
\end{eqnarray}
The elements of $\bar{y}$ correspond to the canonical conjugate operators 
to those of Eq.~\eqref{measurement}; 
i.e. the CCR $y(t)\bar{y}^\top(t') - (\bar{y}(t')y^\top(t))^\top=i\delta(t-t')I$ 
holds.

If we want to measure all the quadratures of $\hat{\cal W}^{\rm out}$, 
it is still possible by introducing additional noise fields 
$\hat{\cal V}=[\hat Q'_1, \hat P'_1, \dots, \hat Q'_m, \hat P'_m]^\top$ 
and performing Homodyne measurement on the joint fields composed of 
$\hat{\cal W}^{\rm out}$ and $\hat{\cal V}$; 
that is, the output equation is given by 
\begin{equation}
\label{heterodyne measurement}
   y = M_1 
        \left[ \begin{array}{c} 
                 \hat{\cal W}^{\rm out} \\
                 \hat{\cal V} \\
               \end{array}\right] 
      = M_1 \left[ \begin{array}{c} 
                 C \\
                 0 \\
               \end{array}\right] \hat x 
      + M_1 \left[ \begin{array}{c} 
                    \hat{\cal W} \\
                    \hat{\cal V} \\
                 \end{array}\right], 
\end{equation}
where in this case $M_1$ is with the size $2m\times 4m$ and it satisfies 
$M_1 \Sigma_{2m} M_1^\top=0$, etc. 
We thus have $2m$ measurement outcomes, though they are subjected to 
the additional noise. 
Note that, by simply replacing $C$ and $\hat{\cal W}$ by $[C^\top, 0]^\top$ 
and $[\hat{\cal W}^\top, \hat{\cal V}^\top]^\top$,  this dual 
Homodyne detection scheme can be represented by Eqs.~\eqref{linear dynamics} 
and \eqref{measurement}. 
Hence in what follows, without loss of generality, we use 
Eq.~\eqref{measurement} to represent the most general linear measurement.

%%%%%%%%%%%%%%%%%%%%%%%%%%%%%%%%%%%%%%%%%%%

\subsection{Examples}

\begin{figure}[t]
\centering 
\includegraphics[width=8.5cm]{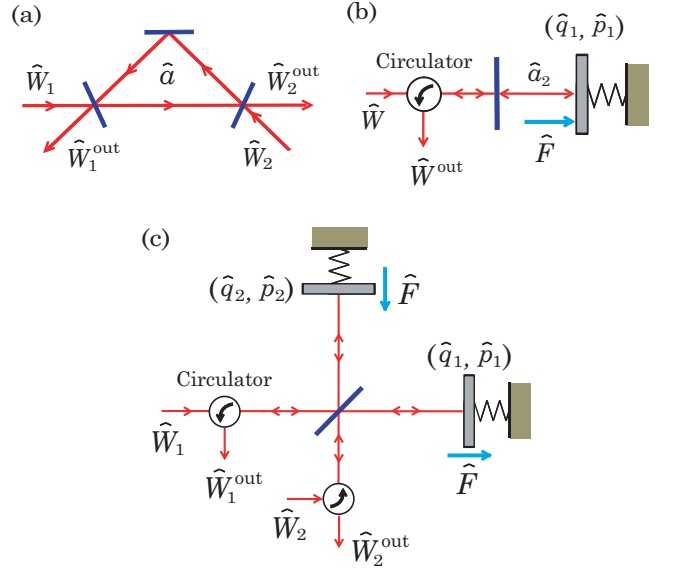}
\caption{
Examples of open linear quantum systems. 
(a) Optical cavity with two input and output fields, 
(b) Mechanical oscillator coupled to an optical cavity, and 
(c) Michelson's interferometer with two identical oscillators. 
}
\label{examples}
\end{figure}

{\bf (i)} 
A simple open linear system is an empty optical cavity with two input 
and output fields, depicted in Fig.~\ref{examples}~(a). 
The system equations are given by 
\begin{eqnarray*}
& & \hspace*{0em}
    \frac{d \hat a}{dt} = -(\kappa_1 + \kappa_2) \hat a
         - \sqrt{2\kappa_1} \hat A_1 - \sqrt{2\kappa_2} \hat A_2,
\nonumber \\ & & \hspace*{0em}
    \hat A_1^{\rm out} =\sqrt{2\kappa_1} \hat a + \hat A_1,~~ 
    \hat A_2^{\rm out} =\sqrt{2\kappa_2} \hat a + \hat A_2. 
\end{eqnarray*}
$\hat a$ is the annihilation operator of the cavity mode. 
$\hat A_j$ and $\hat A_j^{\rm out}$ are the white noise operators of 
the $j$th incoming and the outgoing optical fields, respectively. 
$\kappa_j$ is the coupling strength between $\hat a$ and the $j$th 
field, which is proportional to the transmissivity of the coupling mirror. 
In this paper we express the variables in the quadrature form, which 
in this case are defined as $\hat x =[\hat q, \hat p]^\top$ with 
$\hat q=(\hat a + \hat a^*)/\sqrt{2}$ and 
$\hat p=(\hat a - \hat a^*)/\sqrt{2}i$. 
Also $\hat{W}_j = [\hat Q_j, \hat P_j]^\top$ with the field quadratures 
\eqref{noise quadrature}. 
Then, the above system equations are rewritten as 
\begin{eqnarray*}
& & \hspace*{0em}
    \frac{d\hat x}{dt} = -(\kappa_1+\kappa_2) \hat x 
             - \sqrt{2\kappa_1} \hat{W}_1
             - \sqrt{2\kappa_2} \hat{W}_2,~~~~
\nonumber \\ & & \hspace*{0em}
    \hat{W}_1^{\rm out}=\sqrt{2\kappa_1}\hat x + \hat{W}_1,~~~
    \hat{W}_2^{\rm out}=\sqrt{2\kappa_2}\hat x + \hat{W}_2.
\end{eqnarray*}
Typically this system works as a low-pass filter \cite{BachorBook}; 
that is, for the noisy input field $\hat{W}_1$, the corresponding 
mode-cleaned output field $\hat{W}_2^{\rm out}$ is generated, 
which will be used later for e.g. some quantum information processing. 
To attain this goal, $\hat{W}_1^{\rm out}$ is measured to detect 
the error signal for locking the optical path length in the cavity. 
Note that $\hat{W}_2$ is a vacuum field. 
That is, in this case, the two input-output fields have different roles.

{\bf (ii)} 
The mechanical oscillator shown in Fig.~\ref{examples} (b) can also be 
modeled as a linear system. 
This system is composed of a mechanical oscillator with mode 
$(\hat q_1, \hat p_1)$ and a cavity with mode 
$\hat a_2=(\hat q_2+i\hat p_2)/\sqrt{2}$. 
The cavity couples to a probe field $\hat{W} = [\hat Q, \hat P]^\top$. 
After linearization, the system equation of 
$\hat x =[\hat q_1, \hat p_1, \hat q_2, \hat p_2]^\top$ is obtained as
\begin{eqnarray*}
& & \hspace*{-1em}
    \frac{d\hat x}{dt}
     =\left[ \begin{array}{cc|cc} 
                 0 & 1/m & 0 & 0 \\
                 -m\omega^2 & 0 & \kappa & 0 \\ \hline
                 0 & 0 & -\gamma & 0 \\
                 \kappa & 0 & 0 & -\gamma 
               \end{array}\right]
      \hat x 
     - \sqrt{2\gamma}
        \left[ \begin{array}{cc} 
                 0 & 0 \\
                 0 & 0 \\ \hline
                 1 & 0 \\
                 0 & 1 \\
               \end{array}\right]\hat{W}, 
\nonumber \\ & & \hspace*{-1em}
    \hat{W}^{\rm out}
      =\sqrt{2\gamma}
        \left[ \begin{array}{cc|cc} 
                 0 & 0 & 1 & 0 \\
                 0 & 0 & 0 & 1 \\ 
               \end{array}\right]
                 \hat x + \hat{W}. 
\end{eqnarray*}
$m$ and $\omega$ are the mass and the resonant frequency of the 
oscillator. 
$\kappa$ is the coupling constant between the oscillator and the 
cavity field, which is proportional to the strength of radiation pressure 
force. 
$\gamma$ is the coupling constant between the cavity and the probe field. 
As indicated from the equations, it is possible to extract some information 
about the oscillator's behavior by measuring the probe output field 
$\hat{W}^{\rm out}$. 
A typical situation is that the oscillator is pushed by an external force 
$\hat F$ with unknown strength; 
we attempt to estimate this value, by measuring $\hat{W}^{\rm out}$. 
The oscillator's motion is usually much slower than that of the cavity field, 
thus we can adiabatically eliminate the cavity mode and have a reduced 
dynamical equation of only the oscillator: 
\begin{eqnarray}
& & \hspace*{0em}
\label{reduced dynamics oscillator}
     \frac{d\hat x}{dt}
     =\left[ \begin{array}{cc} 
                 0 & 1/m  \\
                 -m\omega^2 & 0 \\
             \end{array}\right]
      \hat x 
     + \sqrt{\lambda}
        \left[ \begin{array}{c} 
                 0  \\
                 1  \\
               \end{array}\right]\hat{Q}
     + \left[ \begin{array}{c} 
                 0  \\
                 1  \\
               \end{array}\right]\hat{F},
\nonumber \\ & & \hspace*{0em}
     \hat{W}^{\rm out}
      = \left[ \begin{array}{c} 
                 \hat{Q}^{\rm out} \\
                 \hat{P}^{\rm out} \\ 
               \end{array}\right]
      =\sqrt{\lambda}
        \left[ \begin{array}{cc} 
                 0 & 0 \\
                 1 & 0 \\ 
               \end{array}\right]
                 \hat x 
       + \left[ \begin{array}{c} 
                 \hat{Q} \\
                 \hat{P} \\ 
               \end{array}\right], 
\end{eqnarray}
where $\lambda=2\kappa^2/\gamma$ represents the strength of the 
direct coupling between the oscillator and the probe field. 
This equation clearly shows that only $\hat{P}^{\rm out}$ contains the 
information about the oscillator and accordingly $\hat F$; 
thus $\hat{P}^{\rm out}$ should be measured, implying $M_1=[0, 1]$ 
in Eq.~\eqref{measurement}.

{\bf (iii)} 
The last example is the Michelson's interferometer composed of two identical 
mechanical oscillators with mass $m$ and resonant frequency $\omega$, 
depictd in Fig.~\ref{examples}~(c). 
This is a simplest configuration among various schemes that are expected 
to have capability of direct detection of a gravitational wave (GW) 
\cite{BraginskyBook,Caves1980,Chen2013,MiaoThesis}. 
A basic detection mechanism is as follows. 
A coherent light field $\hat{W}_1$ is injected into the left input port 
(bright port), while in the other port (dark port) the input $\hat{W}_2$ 
is set to be a vacuum. 
If a gravitational wave comes, one arm shrinks while the other one extends, 
thereby the oscillators experience tiny force along opposite directions, 
$\hat{F}$ and $-\hat{F}$. 
As a result the dynamics of the two oscillators can be modeled by the 
combination of Eq.~\eqref{reduced dynamics oscillator}:
\begin{eqnarray}
\label{GW without control}
& & \hspace*{-1em}
    \frac{d\hat x}{dt}
     =\left[ \begin{array}{cc|cc} 
       & 1/m & & \\
      -m\omega^2 & & & \\ \hline
       & & & 1/m \\
       & & -m\omega^2 & \\
               \end{array}\right]
      \hat x 
\nonumber \\ & & \hspace*{2em}
   \mbox{}
     + \left[ \begin{array}{c} 
                 0  \\
                 \sqrt{\lambda} \\ \hline
                 0 \\
                 \sqrt{\lambda} \\
               \end{array}\right]\hat Q_1
     + \left[ \begin{array}{c} 
                 0 \\
                 \sqrt{\lambda} \\ \hline
                 0 \\
                 -\sqrt{\lambda} \\
               \end{array}\right]\hat Q_2
     + \left[ \begin{array}{c} 
                 0 \\
                 1 \\ \hline
                 0 \\
                 -1 \\
               \end{array}\right]\hat F, 
\nonumber \\ & & \hspace*{-1em}
    \left[ \begin{array}{c} 
                 \hat{W}_1^{\rm out} \\
                 \hat{W}_2^{\rm out}
               \end{array}\right]
      =\sqrt{\lambda}
        \left[ \begin{array}{cc|cc} 
           & 0 & & 0 \\
           1 & & 1 & \\ \hline
           & 0 & & 0 \\
           1 & & -1 & \\
        \end{array}\right]
                 \hat x 
           + \left[ \begin{array}{c} 
                 \hat{W}_1 \\
                 \hat{W}_2
               \end{array}\right]. 
\end{eqnarray}
Let us rewrite this equation in terms of the common modes 
$\hat q_1'=(\hat q_1+\hat q_2)/\sqrt{2},~
\hat p_1'=(\hat p_1+\hat p_2)/\sqrt{2}$ and the differential modes 
$\hat q_2'=(\hat q_1-\hat q_2)/\sqrt{2},~
\hat p_2'=(\hat p_1-\hat p_2)/\sqrt{2}$. 
Then these two modes are decoupled and the force $\hat F$ appears 
only in the dynamics of $\hat x_2'=[\hat q_2', \hat p_2']^\top$, which 
is exactly the same as Eq.~\eqref{reduced dynamics oscillator}: 
\begin{eqnarray}
\label{GW without control trans}
& & \hspace*{-2em}
    \frac{d\hat x_2'}{dt}
     =\left[ \begin{array}{cc} 
                 0 & 1/m  \\
                 -m\omega^2 & 0 \\
             \end{array}\right]
      \hat x_2' 
     + \sqrt{\lambda}
        \left[ \begin{array}{c} 
                 0  \\
                 1  \\
               \end{array}\right]\hat{Q}_2
     + \left[ \begin{array}{c} 
                 0  \\
                 1  \\
               \end{array}\right]\hat{F},
\nonumber \\ & & \hspace*{-2em}
     \hat{Q}_2^{\rm out}=\hat Q_2,~~~
     \hat{P}_2^{\rm out}=\sqrt{\lambda}\hat q_2' + \hat P_2. 
\end{eqnarray}
Thus, ideally, by measuring $\hat{P}_2^{\rm out}$ we can detect 
$\hat F$.

%%%%%%%%%%%%%%%%%%%%%%%%%%%%%%%%%%%%%%%%%%%%%
%%%%%%%%%%%%%%%%%%%%%%%%%%%%%%%%%%%%%%%%%%%%%
%%%%%%%%%%%%%%%%%%%%%%%%%%%%%%%%%%%%%%%%%%%%%

\section{System theoretic characterization of BAE, QND, DFS}

The problem considered in this paper is to design a MF/CF controller 
connected to the plant system so that the plant or the whole closed-loop 
system achieves a certain control goal. 
We consider the following three goals: 
realization of back-action evasion (BAE) measurement, generation of 
a quantum non-demolished (QND) variable, and generation of a 
decoherence-free subsystem (DFS). 
Actually there are a lot of works investigating their mathematical 
characterizations, physical realizations, and applications especially in 
quantum information science. 
This section shows system theoretic characterizations of these notions 
in terms of controllability and observability properties or transfer 
functions, in a consistent way with the standard definitions.

%%%%%%%%%%%%%%%%%%%%%%%%%%%%%%%%%%%%%%%%%%%%

\subsection{BAE}

The idea of BAE originally comes from the research for GW detection. 
The Michelson's interferometer described in Sec.~II-C is a simplest 
system for this purpose, and we now know from 
Eq.~\eqref{GW without control trans} that the measurement output 
$y=\hat{P}_2^{\rm out}=\sqrt{\lambda}\hat q_2' + \hat P_2$ would 
offer some information about $\hat F$. 
The issue is that, in addition to the unavoidable noise $\hat P_2$ called 
the {\it shot noise}, the output $y$ contains the conjugate $\hat Q_2$, 
which is called the {\it back-action (BA) noise}, as seen explicitly in the 
Laplace domain: 
\[
    y[s] = \frac{\sqrt{\lambda}}{m(s^2+\omega^2)}
         \big( \sqrt{\lambda}\hat{Q}_2[s] + \sqrt{2}\hat{F}[s] \big) 
             + \hat{P}_2[s]. 
\]
The slight change of the oscillator's position due to the GW effect, $\hat g$, 
is defined in the Fourier domain $s=i\Omega$ as 
$\hat{F}[i\Omega]=-mL\Omega^2 \hat{g}[i\Omega]$, where $L$ is the 
optical path length in the interferometer. 
Hence under the assumption $\Omega \gg \omega$, the normalized signal 
containing $\hat g$ is given by 
\[
   \tilde{y}[i\Omega]
    =\frac{y[i\Omega]}{2\sqrt{\lambda}L}
    = \hat{g}[i\Omega] 
           + \frac{\sqrt{\lambda}}{mL\Omega^2}\hat{Q}_2[i\Omega] 
                + \frac{1}{2\sqrt{\lambda}L}\hat{P}_2[i\Omega].
\]
The noise power of $\tilde{y}$ is bounded from below by the following 
{\it standard quantum limit (SQL)}: 
\begin{eqnarray}
\label{SQL}
& & \hspace*{-2.5em}
   S[i\Omega]=\mean{|\tilde{y}-\hat{g}|^2}
      = \frac{\lambda}{m^2L^2\Omega^4}\mean{|\hat{Q}_2|^2} 
                + \frac{1}{4\lambda L^2}\mean{|\hat{P}_2|^2}
\nonumber \\ & & \hspace*{0em}
      \geq 2\sqrt{ \frac{\mean{|\hat{Q}_2|^2}\mean{|\hat{P}_2|^2}}
                        {4m^2 L^4 \Omega^4} }
      \geq \frac{1}{2mL^2 \Omega^2} = S_{\rm SQL}[i\Omega].
\end{eqnarray}
The last inequality is due to the Heisenberg uncertainty relation 
$\mean{|\hat{Q}_2|^2}\mean{|\hat{P}_2|^2} \geq 1/4$. 
(For the simple notation, the power spectrum is defined without involving the 
delta function.) 
The SQL appears because the output $y$ contains the BA noise $\hat Q_2$ 
in addition to the shot noise $\hat P_2$. 
Thus, towards high-precision detection of $\hat g$, a special system 
configuration should be devised so that $y$ is free from $\hat Q_2$. 
That is, we need BAE. 
In fact, if BAE is realized, then by injecting a $\hat P_2$-squeezed light 
field into the dark port, we can possibly reduce the noise power below the SQL 
and may have chance to detect $\hat g$; 
for some specific configurations achieving BAE, see 
\cite{BraginskyBook,Caves1980,Tsang2010,Chen2013,MiaoThesis}.

The above discussion can be generalized for the system \eqref{linear dynamics} 
and \eqref{measurement}. 
Let us assume that the signal to be detected is contained in the output 
\eqref{measurement}: 
\begin{equation}
\label{BAE measurement}
      y = M_1 \hat{\cal W}^{\rm out} 
        = M_1 C \hat x + M_1 \hat{\cal W}
        = M_1 C \hat x + \hat{\cal Q}.
\end{equation}
Hence, $\hat{\cal Q}=M_1 \hat{\cal W}$ is the shot noise, which must 
appear in $y$. 
The BA noise is then given by the conjugate $\hat{\cal P}=M_2 \hat{\cal W}$. 
Note that these are vectors of operators: 
$\hat{\cal Q}=[\hat Q_1, \ldots, \hat Q_m]^\top$ and 
$\hat{\cal P}=[\hat P_1, \ldots, \hat P_m]^\top$. 
The matrices $M_1$ and $M_2$ satisfy several conditions \eqref{M1 M2 conditions}; 
in particular $M_1^\top M_1+M_2^\top M_2=I$ holds and leads to 
$\hat{\cal W}=M_1^\top\hat{\cal Q}+M_2^\top\hat{\cal P}$. 
Hence Eq.~\eqref{linear dynamics} is rewritten as 
\begin{equation}
\label{BAE system}
\frac{d\hat x}{dt} = A\hat x 
       + \Sigma_n C^\top \Sigma_m M_1^\top\hat{\cal Q} 
       + \Sigma_n C^\top \Sigma_m M_2^\top\hat{\cal P}. 
\end{equation}
BAE is realized, if the output \eqref{BAE measurement} does not contain 
the BA noise $\hat{\cal P}$. 
(We will not consider the so-called variational measurement approach, in 
which case $M_1$ is frequency dependent.) 
In the language of linear systems theory, as stated in Eq.~\eqref{classical BAE}, 
this condition means that there is no subsystem that is controllable w.r.t. 
$\hat{\cal P}$ and observable w.r.t. $y$; i.e. 
\begin{equation}
\label{def of BAE alg}
    \mbox{{\bf BAE:}}~~~ 
    {\rm Ker}({\cal C}_{\hat{\cal P}}^\top)^c \cap {\rm Range}({\cal O}_y^\top) 
       = \emptyset, 
\end{equation}
where ${\cal C}_{\hat{\cal P}}$ is the controllability matrix generated from 
$(A, \Sigma_n C^\top \Sigma_m M_2^\top)$ and ${\cal O}_y$ is the 
observability matrix generated from $(A, M_1C)$. 
Further, again as described in Eq.~\eqref{classical BAE}, the condition 
\eqref{def of BAE alg} is equivalent to 
\begin{equation}
\label{def of BAE alg 2}
         M_1C A^k \Sigma_n C^\top \Sigma_m M_2^\top=0,~~\forall k\geq 0. 
\end{equation}
Under this condition, the system equations \eqref{BAE measurement} and 
\eqref{BAE system} are represented in a transformed coordinate by
\begin{eqnarray*}
& & \hspace*{-1.1em}
     \frac{d}{dt}
      \left[ \begin{array}{c} 
                 \hat x'_1 \\
                 \hat x'_2 \\
               \end{array}\right]
     =\left[ \begin{array}{cc} 
                 A_{11} & 0 \\
                 A_{21} & A_{22} \\
               \end{array}\right]
       \left[ \begin{array}{c} 
                 \hat x'_1 \\
                 \hat x'_2 \\
               \end{array}\right]
       + \left[ \begin{array}{c} 
                 B_{11} \\
                 B_{21} \\
               \end{array}\right]\hat{\cal Q}
       + \left[ \begin{array}{c} 
                 0 \\
                 B_{22} \\
               \end{array}\right]\hat{\cal P},
\nonumber \\ & & \hspace*{-1em}
      y=[C_1,~0]
       \left[ \begin{array}{c} 
                 \hat x'_1 \\
                 \hat x'_2 \\
               \end{array}\right] + \hat{\cal Q}, 
\end{eqnarray*}
showing that actually there is no signal flow from $\hat{\cal P}$ to $y$. 
It is also obvious from this equation that, similar to the classical case 
\eqref{classical BAE transfer fn}, the equivalent characterization to 
Eq.~\eqref{def of BAE alg} in terms of the transfer function is given by 
\begin{equation}
\label{def of BAE}
          \mbox{{\bf BAE:}}~~~ 
          \Xi_{\hat{\cal P}\rightarrow y}[s]=0,~~\forall s.
\end{equation}
Finally, note that achieving the above BAE condition \eqref{def of BAE alg} 
or \eqref{def of BAE} itself does not necessarily mean the improvement of 
signal sensitivity; 
actually in the case of GW force sensing discussed in Sec.~VII-B, we need 
squeezing of the input field in addition to the BAE property for realizing 
such operational improvement.

%%%%%%%%%%%%%%%%%%%%%%%%%%%%%%%%%%%%%%%%%%%

\subsection{QND}

\begin{figure}[t]
\centering 
\includegraphics[width=6cm]{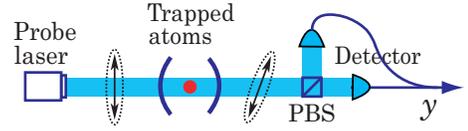}
\caption{
Atomic ensemble under continuous measurement via Faraday rotation. 
PBS: polarized beam splitter. 
}
\label{Spin QND}
\end{figure}

Next to see the idea of QND variables, let us here study the atomic ensemble 
trapped in a cavity \cite{WisemanBook,Takahashi,Stockton,BoutenPRA07} 
shown in Fig.~\ref{Spin QND}. 
The atoms couple with a probe polarized light field, via the Faraday 
interaction. 
In terms of the total energy operator $\hat J_z$ and its conjugates 
$\hat J_x$ and $\hat J_y$, which satisfy the CCRs e.g. 
$\hat J_y\hat J_z - \hat J_z\hat J_y = i\hat J_x$, the ideal dynamics of 
atomic ensemble is described by 
\begin{equation}
\label{spin ensemble QSDE}
     \frac{d}{dt}
     \left[ \begin{array}{c} 
                 \hat J_x \\
                 \hat J_y \\
                 \hat J_z \\
               \end{array}\right]
      =\left[ \begin{array}{ccc} 
                 -M/2 & -\sqrt{2M}\hat P & 0 \\
                 \sqrt{2M}\hat P & -M/2 & 0 \\
                 0 & 0 & 0 \\
               \end{array}\right]
         \left[ \begin{array}{c} 
                 \hat J_x \\
                 \hat J_y \\
                 \hat J_z \\
               \end{array}\right]. 
\end{equation}
$\hat P$ is the phase quadrature of the input field's noise operator 
corresponding to the polarization, and $M$ represents the coupling 
strength between the atoms and the field. 
In this setting, the amplitude quadrature of the output field should 
be measured, giving the following measurement output equation: 
\[
      y = \hat Q^{\rm out} = \sqrt{2M}\hat J_z + \hat Q. 
\]
From these two equations, we find that, through the Faraday interaction, 
the polarization of the probe field rotates depending on the total energy 
$\hat J_z$, but $\hat J_z$ itself does not change; 
that is, $\hat J_z$ is a QND variable that can be measured without being 
disturbed. 
Typically $M$ is relatively small, and then the system variables obey a 
skew-Hermitian dynamics, implying that they preserve 
$\hat J_x^2+\hat J_y^2+\hat J_z^2$. 
Hence, in the large ensemble limit and in the short time period, the 
dynamics is constrained in the tangent space of this super-sphere 
with radius $J=N/2$ ($N$ is the number of atoms). 
In particular let us set $\hat J_x$ to be a constant $J$ rather than 
the operator-valued variable. 
Then the system variables are given by the usual CCR pairs 
$\hat q=\hat J_y/\sqrt{J}$ and $\hat p=\hat J_z/\sqrt{J}$ satisfying 
$\hat q\hat p-\hat p\hat q=i$, and the above system dynamics can be 
simplified to the following linear equation: 
\[
    \frac{d}{dt}
       \left[ \begin{array}{c} 
                 \hat q \\
                 \hat p \\
               \end{array}\right]
       = \sqrt{\mu} 
           \left[ \begin{array}{c} 
                   1 \\
                   0 \\
                  \end{array}\right]\hat P,~~~
      y = \sqrt{\mu} \hat p + \hat Q,
\]
where $\mu=2MJ$. 
Clearly, $\hat p$ is not disturbed by the noise while it appears in the output 
signal, thus $\hat p$ is a QND variable. 
A merit of QND measurement is in the application to state preparation; 
if a QND variable exists, it is sometimes possible to deterministically 
stabilize its eigenstate by feedback \cite{WisemanBook}, which can be 
highly non-classical such as a spin-squeezed state \cite{Takahashi,Stockton}.

As in the BAE case, we have a general characterization of the linear 
system \eqref{linear dynamics} and \eqref{measurement} having 
a QND variable. 
Let $\hat r=v^\top \hat x$ be a QND variable with $v\in{\mathbb R}^{2n}$. 
Then, by definition, $\hat r$ must not be affected by the input field 
$\hat{\cal W}$, while it appears in the output signal \eqref{measurement}, 
$y=MC\hat x + M\hat{\cal W}$. 
This means that, in the language of linear systems theory, 
$\hat r=v^\top \hat x$ is uncontrollable w.r.t. $\hat{\cal W}$ and 
observable w.r.t. $y$. 
Thus, the iff condition for a QND variable to exist is given by 
\begin{equation}
\label{def of QND alg}
    \mbox{{\bf QND:}}~~~ 
    {\rm Ker}({\cal C}_{\hat{\cal W}}^\top) \cap 
    {\rm Range}({\cal O}_y^\top) \neq \emptyset, 
\end{equation}
and the vector $v$ lives in this intersection. 
Here, ${\cal C}_{\hat{\cal W}}$ and ${\cal O}_y$ are the controllability 
and observability matrices of the system \eqref{linear dynamics} and 
\eqref{measurement}. 
Note that the condition $v\in{\rm Ker}({\cal C}_{\hat{\cal W}}^\top)$ 
can be explicitly represented by 
\begin{equation}
\label{def of QND alg 2}
         v^\top A^k \Sigma_n C^\top = 0,~~\forall k\geq 0. 
\end{equation}
Now let us collect QND variables into a single vector $\hat x_2'$. 
Then, as described in Sec.~II-A, $\hat x_2'$ constitutes an uncontrollable 
subsystem w.r.t. $\hat{\cal W}$, which can be clearly seen in the 
transformed coordinate: 
\begin{eqnarray*}
& & \hspace*{0em}
     \frac{d}{dt}
      \left[ \begin{array}{c} 
                 \hat x_1' \\
                 \hat x_2' \\
               \end{array}\right]
     =\left[ \begin{array}{cc} 
                 A_{11} & A_{12} \\
                 0 & A_{22} \\
               \end{array}\right]
       \left[ \begin{array}{c} 
                 \hat x_1' \\
                 \hat x_2' \\
               \end{array}\right]
       + \left[ \begin{array}{c} 
                 B_1 \\
                 0       \\
               \end{array}\right]\hat{\cal W},~~~
\nonumber \\ & & \hspace*{0em}
      y=[C_1,~C_2]
       \left[ \begin{array}{c} 
                 \hat x_1' \\
                 \hat x_2' \\
               \end{array}\right] + M\hat{\cal W}. 
\end{eqnarray*}
Note $C_2\neq 0$ due to the observability condition. 
Hence, $\hat x'_2$ is free from $\hat{\cal W}$, while it appears in $y$. 
Remarkably, $\hat x'_2$ obeys the closed dynamics 
$d\hat x_2'/dt=A_{22}\hat x_2'$; thus $\hat x'_2$ is a generalization of a 
standard QND variable, which is usually considered to be 
static (i.e. $\hat x_2'(t)=x_2'(0)$, $\forall t$); see \cite{Tsang2012} for 
further detailed discussion. 
The above equation now enables us to obtain the equivalent condition to 
Eq.~\eqref{def of QND alg} in terms of the transfer functions: 
\begin{equation}
\label{def of QND transfer}
    \mbox{{\bf QND:}}~~~ 
    \Xi_{\hat{\cal W} \rightarrow \hat x_2'}[s]=0,~\forall s~~~\&~~~
    \Xi_{\hat x_2' \rightarrow y}[s]\neq 0,~\exists s. 
\end{equation}
%

%%%%%%%%%%%%%%%%%%%%%%%%%%%%%%%%%%%%%%%%%%%

\subsection{DFS}

The idea of the third control goal, generation of a DFS, can be clearly seen 
from the work \cite{Sorensen2007}, which studies a quantum memory 
served by an atomic ensemble in a cavity. 
Each atom has $\Lambda$-type energy levels, constituted by two metastable 
ground states $(\ket{s}, \ket{g})$ and an excited state $\ket{e}$. 
The state transition between $\ket{e}$ and $\ket{g}$ is naturally coupled 
to the cavity mode $\hat a_1$ with strength $g\sqrt{N}$ ($N$ denotes the 
number of atoms), while the $\ket{s}\leftrightarrow\ket{e}$ transition is 
induced by a classical magnetic field with time-varying Rabi frequency 
$\omega(t)$. 
The system variables are the polarization operator 
$\hat a_2=\hat \sigma_{ge}/\sqrt{N}$ and the spin-wave operator
$\hat a_3=\hat \sigma_{gs}/\sqrt{N}$, where $\hat \sigma_{\bullet}$ is 
the collective lowering operator; 
in a large ensemble limit, they can be well approximated by annihilation 
operators. 
Consequently the system dynamics is given by 
\begin{eqnarray}
\label{memory example}
& & \hspace*{-1.2em}
      \frac{d}{dt}
       \left[ \begin{array}{c}
                  \hat{a}_1 \\
                  \hat{a}_2 \\
                  \hat{a}_3 \\
             \end{array} \right]
        =   \left[ \begin{array}{ccc}
                  -\kappa & ig\sqrt{N} & 0   \\
                  ig\sqrt{N} & -i\delta & i\omega   \\
                  0 & i\omega^* & 0 \\
             \end{array} \right]
              \left[ \begin{array}{c}
                   \hat{a}_1 \\
                   \hat{a}_2 \\
                   \hat{a}_3 \\
              \end{array} \right]
          - \left[ \begin{array}{c}
                  \sqrt{2\kappa}  \\
                  0 \\
                  0 \\
             \end{array} \right]
                   \hat A, 
\nonumber \\ & & \hspace*{-1.2em}
      \hat{A}^{\rm out} = \sqrt{2\kappa}\hat{a}_1 + \hat A,
\end{eqnarray}
where $\kappa$ denotes the cavity decay rate and $\delta$ is the detuning 
between the cavity center frequency and the $\ket{s}\leftrightarrow\ket{e}$ 
transition frequency. 
This system works as a quantum memory in the following way. 
First, a state to be stored is carried by an appropriately shaped optical pulse 
on the input field $\hat A$, and it is transferred to the metastable 
state $\ket{s}$; 
the Rabi frequency $\omega(t)$ is suitably designed throughout this writing 
process. 
In the storage stage, the classical magnetic field is turned off, i.e. $\omega(t)=0$. 
It is seen from Eq.~\eqref{memory example} that the spin-wave operator 
$\hat a_3$ is then completely decoupled from the fields $\hat A$ and 
$\hat A^{\rm out}$; 
that is, $\hat a_3$ constitutes a linear DFS, and ideally its state is 
perfectly preserved. 
In the language of systems theory, this DFS is uncontrollable w.r.t. $\hat A$ 
and unobservable w.r.t. $\hat A^{\rm out}$. 
Note that $\hat a_3$ is not a variable on the so-called decoherence-free 
{\it subspace}, which though has the same abbreviation. 
In general, if the system's Hilbert space can be decomposed to 
$({\cal H}_1\otimes {\cal H}_2)\oplus{\cal H}_3$ and ${\cal H}_1$ is 
free from external noise, then it is called the DF subsystem and 
particularly when ${\rm dim}{\cal H}_2=1$ it is called the DF 
subspace \cite{ZanardiPRL1997,Lidar2003}; 
now we are dealing with the case where $\hat a_3$ and 
$(\hat a_1, \hat a_2)$ live in ${\cal H}_1$ and ${\cal H}_2$, respectively, 
while ${\rm dim}{\cal H}_3=0$. 
For other examples of such an infinite dimensional DFS, see 
\cite{ClerkNJP2012,ClerkPRL2012,WangScience,YamamotoArxiv2014,
PrauznerJPA2004,ZambriniPRA2013,YamamotoArxiv2013}.

The above fact reasonably leads to a general characterization of the 
system \eqref{linear dynamics} and \eqref{linear output} that contains 
a DFS. 
By definition, a DFS is completely decoupled from the probe/environment 
field, so it is not affected by $\hat{\cal W}$ and also it does not appear in 
$\hat{\cal W}^{\rm out}$. 
In the language of systems theory, a variable contained in the DFS is 
uncontrollable w.r.t. $\hat{\cal W}$ and unobservable w.r.t. 
$\hat{\cal W}^{\rm out}$. 
Thus the iff condition for a DFS to exist is given by 
\begin{equation}
\label{def of DFS alg}
    \mbox{{\bf DFS:}}~~~ 
    {\rm Ker}({\cal C}_{\hat{\cal W}}^\top) \cap 
    {\rm Range}({\cal O}_{\hat{\cal W}^{\rm out}}^\top)^c \neq \emptyset, 
\end{equation}
where ${\cal C}_{\hat{\cal W}}$ and ${\cal O}_{\hat{\cal W}^{\rm out}}$ 
are the controllability and observability matrices of the system 
\eqref{linear dynamics} and \eqref{linear output}. 
In particular, as seen in Eqs.~\eqref{classical QND} and \eqref{classical QND 2}, 
there always exists a coordinate transformation such that 
$\hat r=v^\top \hat x$ is a variable of the DFS iff the vector 
$v\in{\mathbb R}^{2n}$ is contained in the intersection 
${\rm Ker}({\cal C}_{\hat{\cal W}}^\top) \cap 
{\rm Ker}({\cal O}_{\hat{\cal W}^{\rm out}})$; 
that is, it satisfies 
\begin{equation}
\label{def of DFS alg 2}
         v^\top A^k \Sigma_n C^\top \Sigma_m = 0,~~~
         C A^k v = 0,~~\forall k\geq 0. 
\end{equation}
(A convenient method to construct such $v$ is given in 
\cite{YamamotoArxiv2013}.) 
Then, as in the QND case, by collecting all variables in the DFS into 
a single vector $\hat x_2'$, we find that the system equations can be 
transformed to 
\begin{eqnarray*}
& & \hspace*{0em}
     \frac{d}{dt}
      \left[ \begin{array}{c} 
                 \hat x_1' \\
                 \hat x_2' \\
               \end{array}\right]
     =\left[ \begin{array}{cc} 
                 A_{11} & 0 \\
                 0 & A_{22} \\
               \end{array}\right]
       \left[ \begin{array}{c} 
                 \hat x_1' \\
                 \hat x_2' \\
               \end{array}\right]
       + \left[ \begin{array}{c} 
                 B_1 \\
                 0 \\
               \end{array}\right]\hat{\cal W},~~~
\nonumber \\ & & \hspace*{0em}
      \hat{\cal W}^{\rm out}
        =[C_1,~0]
         \left[ \begin{array}{c} 
                 \hat x_1' \\
                 \hat x_2' \\
                \end{array}\right] + \hat{\cal W}. 
\end{eqnarray*}
Thus $\hat x'_2$ obeys the closed-dynamics $d\hat x_2'/dt=A_{22}\hat x_2'$; 
especially if $A_{22}=0$, the state of $\hat x_2'$ is kept unchanged, and 
the DFS works as a memory. 
Lastly, the condition for $\hat x'_2$ to be a variable in the DFS is given 
in terms of the transfer functions by 
\begin{equation}
\label{def of DFS}
     \mbox{{\bf DFS:}}~~~ 
     \Xi_{\hat{\cal W} \rightarrow \hat x_2'}[s]=0,~~~
     \Xi_{\hat x_2' \rightarrow \hat{\cal W}^{\rm out}}[s]=0,~~
     \forall s.
\end{equation}
Note here again that the condition \eqref{def of DFS alg} or \eqref{def of DFS} 
is only a necessary requirement for the system to have a good memory 
architecture, and it itself does not lead to the improvement of memory 
retrieval fidelity. 
To realize a high-quality quantum memory process, in addition to engineering 
such a DFS, we need a sophisticated method for transferring an input state to 
the memory part. 
For instance by suitable pulse shaping of the input wave packet, lossless 
state transfer to a general linear DFS and accordingly perfect memory 
fidelity can be achieved \cite{YamamotoArxiv2014}.

%%%%%%%%%%%%%%%%%%%%%%%%%%%%%%%%%%%%%%%%%%%
%%%%%%%%%%%%%%%%%%%%%%%%%%%%%%%%%%%%%%%%%%%
%%%%%%%%%%%%%%%%%%%%%%%%%%%%%%%%%%%%%%%%%%%

\section{The no-go theorems: type-1 case}

\begin{figure}[t]
\centering 
\includegraphics[width=6.4cm]{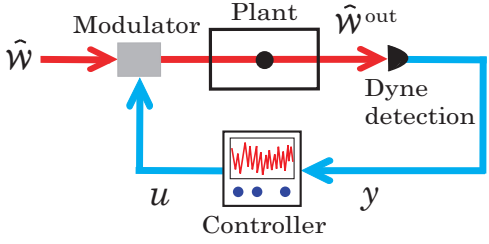}
\caption{
General configuration of the type-1 MF control. 
}
\label{SISO general FB}
\end{figure}

In this paper, we study a general linear system having multi input and 
multi output fields (it is called a MIMO system). 
The first essential question is about which input and output fields should 
be used for feedback. 
We define the {\it type-1 control} as a configuration where at most all the 
input and output fields can be used for this purpose. 
Note that, if the system has single input-output channel such as the one 
shown in Sec.~II-C (ii), the control configuration must be of type-1. 
Figure~\ref{SISO general FB} illustrates the general configuration of 
type-1 MF control. 
That is, at most all the plant's output fields can be measured, and the 
measurement results $y(t)$ are then processed in a classical system 
(controller) that produces a control signal $u(t)$. 
From the standpoint comparing MF and CF, we assume that the control 
is carried out by modulating the input probe fields, which can be 
physically implemented using an electric optical modulator on the 
optical field; 
in the type-1 case, hence, at most all the plant's input fields can be 
modulated using the control signal $u(t)$. 
This section studies the type-1 MF control and shows the no-go theorems 
given in the left column of Table~I.

%%%%%%%%%%%%%%%%%%%%%%%%%%%%%%%%%%%%%%%%%%%

\subsection{The closed-loop system with type-1 MF}

As described above, the MF control is carried out by modulating the input 
probe fields. 
This mathematically means that the input field is replaced by 
$\hat {\cal W} + u$, where $u=[u_1, \ldots, u_{2m}]^\top$ is a 
vector of classical control signals representing the modulation. 
Hence our plant system is now given by 
\begin{eqnarray}
& & \hspace*{0em}
\label{type 1 plant}
   \frac{d\hat x}{dt} = A \hat x 
       + \Sigma_n C^\top \Sigma_m (\hat {\cal W} + u), 
\\ & & \hspace*{0em}
\label{type 1 output}
    \hat {\cal W}^{\rm out} = C\hat x + \hat {\cal W} + u. 
\end{eqnarray}
Note that the output field is directly controlled. 
(In what follows we omit the subscript of $\Sigma_\bullet$ for notational 
simplicity.) 
The output signal is obtained by measuring $\hat {\cal W}^{\rm out}$:
\begin{equation}
\label{type 1 meas output}
   y = M_1\hat {\cal W}^{\rm out} 
      = M_1C\hat x + \hat {\cal Q} + M_1u, 
\end{equation}
where $\hat {\cal Q}=M_1\hat {\cal W}$ with $M_1$ the symplectic matrix 
defined in Sec.~II-B. 
Also the conjugate noise operator is given by $\hat {\cal P}=M_2\hat {\cal W}$; 
these matrices satisfy the conditions \eqref{M1 M2 conditions}.

The controller is a classical system that processes the measurement 
result $y(t)$ and produces the control signal $u(t)$. 
The dynamical equation of this system can be generally represented by 
\begin{equation}
\label{type 1 controller}
    \frac{dx_K}{dt}=A_K x_K + B_K y,~~~~
    u = C_K x_K, 
\end{equation}
where $(A_K, B_K, C_K)$ are the parameter matrices to be designed. 
$x_K$ is the vector of controller's variables, and its dimension is also 
a parameter; 
hence there is a large freedom in engineering the controller. 
Note that the matrices are not necessarily of full rank, meaning that 
in this case some output fields are not measured or some input fields 
are not modulated. 
Combining all the above equations, we have the closed-loop 
(quantum-classical hybrid) dynamics of 
$\hat x_e=[\hat x^\top, x_K^\top]^\top$ as follows; 
\begin{eqnarray}
& & \hspace*{-2em}
\label{type 1 closed-loop dynamics QND}
    \frac{d\hat x_e}{dt}
     =\left[ \begin{array}{cc} 
       A & \Sigma C^\top \Sigma C_K \\
       B_K M_1 C & A_K + B_K M_1 C_K \\
      \end{array}\right] \hat x_e
\nonumber \\ & & \hspace*{5em}
     \mbox{}
     + \left[ \begin{array}{c} 
                 \Sigma C^\top \Sigma \\
                 B_K M_1\\
               \end{array}\right]\hat{\cal W}, 
\\ & & \hspace*{-2em}
\label{type 1 closed-loop output}
    y=[M_1C, M_1C_K] \hat x_e
      + \hat{\cal Q}.
\end{eqnarray}
Hence, $\hat{\cal Q}$ is the shot noise. 
Equation~\eqref{type 1 closed-loop dynamics QND} can be expressed 
in terms of the quadratures $\hat{\cal Q}$ and $\hat{\cal P}$ as: 
\begin{eqnarray}
& & \hspace*{-1em}
\label{type 1 closed-loop dynamics BAE}
    \frac{d\hat x_e}{dt}
     =\left[ \begin{array}{cc} 
       A & \Sigma C^\top \Sigma C_K \\
       B_K M_1 C & A_K + B_K M_1 C_K \\
      \end{array}\right] \hat x_e
\nonumber \\ & & \hspace*{1em}
     + \left[ \begin{array}{c} 
                 \Sigma C^\top \Sigma M_1^\top \\
                 B_K \\
               \end{array}\right]\hat{\cal Q}
     + \left[ \begin{array}{c} 
                 \Sigma C^\top \Sigma M_2^\top \\
                 0 \\
               \end{array}\right]\hat{\cal P}, 
\end{eqnarray}
due to $\hat{\cal W}=M_1^\top\hat{\cal Q} + M_2^\top\hat{\cal P}$. 
We aim to find a set of matrices $(A_K, B_K, C_K, M_1, M_2)$ that 
achieves the control goals described in Sec. III; 
but as shown below, it is impossible to accomplish those tasks.

%%%%%%%%%%%%%%%%%%%%%%%%%%%%%%%%%%%%%%%%%%%

\subsection{BAE}

Suppose that BAE holds for the closed-loop dynamics 
\eqref{type 1 closed-loop dynamics BAE} with output 
\eqref{type 1 closed-loop output}; 
that is, the condition \eqref{def of BAE alg} holds for this system, 
which is now 
${\rm Ker}({\cal C}_{\hat{\cal P}}^\top)^c\cap{\rm Range}({\cal O}_y^\top) 
= \emptyset$. 
(Equivalently, the transfer function of the closed-loop system satisfies 
$\Xi_{\hat{\cal P}\rightarrow y}^{(fb)}[s]=0, \forall s$.) 
This is further equivalent, as implied by Eq.~\eqref{def of BAE alg 2}, to 
\begin{eqnarray}
& & \hspace*{-3em}
    [M_1C, M_1C_K]
      \left[ \begin{array}{cc} 
       A & \Sigma C^\top \Sigma C_K \\
       B_K M_1 C & A_K + B_K M_1 C_K \\
      \end{array}\right]^k
\nonumber \\ & & \hspace*{5em}
       \times 
               \left[ \begin{array}{c} 
                  \Sigma C^\top \Sigma M_2^\top \\
                  0 \\
               \end{array}\right]
               =0,~~\forall k\geq 0.
\end{eqnarray}
First, the case $k=0$ leads to $M_1C \Sigma C^\top \Sigma M_2^\top = 0$. 
Then, using this condition, we find that the case $k=1$ yields 
$M_1C A \Sigma C^\top \Sigma M_2^\top = 0$. 
This further allows us from the case $k=2$ to have 
$M_1C A^2 \Sigma C^\top \Sigma M_2^\top = 0$. 
Repeating the same procedure we eventually obtain 
\[
     M_1C A^k \Sigma C^\top \Sigma M_2^\top = 0,~~\forall k\geq 0. 
\]
This is exactly the BAE condition for the {\it original} plant system 
\eqref{BAE measurement} and \eqref{BAE system}, i.e. 
\[
      \frac{d\hat x}{dt} = A \hat x 
         + \Sigma C^\top \Sigma M_1^\top \hat {\cal Q} 
         + \Sigma C^\top \Sigma M_2^\top \hat {\cal P},~
       y = M_1C\hat x + \hat {\cal Q}.
\]
Equivalently, the transfer function of the original plant system satisfies 
$\Xi_{\hat{\cal P}\rightarrow y}^{(o)}[s]=0,~\forall s$. 
Thus the contrapositive of this result yields the following theorem.

{\bf Theorem 1:} 
If the original plant system does not have the BAE property, 
then, any type-1 MF control cannot realize BAE for the closed-loop 
system.

%%%%%%%%%%%%%%%%%%%%%%%%%%%%%%%%%%%%%%%%%%%

\subsection{QND}

First of all, let us consider the case where the closed-loop system 
\eqref{type 1 closed-loop dynamics QND} and \eqref{type 1 closed-loop output} 
has a QND variable $\hat r$. 
This should be ``purely quantum", meaning that $\hat r$ is composed of 
only the quantum variables 
$\hat x=[\hat q_1, \hat p_1, \ldots, \hat q_n, \hat p_n]^\top$; 
hence it is of the form $\hat r=v^\top \hat x=\tilde{v}^\top\hat x_e$ 
with $\tilde{v}=[v^\top, 0^\top]^\top$. 
As described in Eq.~\eqref{def of QND alg}, this means 
$\tilde{v}\in{\rm Ker}({\cal C}_{\hat{\cal W}}^\top)\cap
{\rm Range}({\cal O}_y^\top)$, with ${\cal C}_{\hat{\cal W}}$ and 
${\cal O}_y$ the controllability and observability matrices of the system 
\eqref{type 1 closed-loop dynamics QND} and \eqref{type 1 closed-loop output}. 
To prove the no-go theorem, the following two facts are useful. 
First, $\tilde{v}\in{\rm Ker}({\cal C}_{\hat{\cal W}}^\top)$ means that 
\[
    [v^\top, 0^\top]
     \left[ \begin{array}{cc} 
       A & \Sigma C^\top \Sigma C_K \\
       B_K M_1 C & A_K + B_K M_1 C_K \\
      \end{array}\right]^k
         \left[ \begin{array}{c} 
             \Sigma C^\top \Sigma \\
             B_K M_1\\
         \end{array}\right] = 0, 
\]
for all $k\geq 0$. 
It follows from a similar procedure as in the BAE case that this is equivalent 
to $v^\top A^k\Sigma C^\top \Sigma=0$, $\forall k\geq 0$; 
i.e. $v\in{\rm Ker}({\cal C}_{\hat{\cal W}}^\top)$ with ${\cal C}_{\hat{\cal W}}$ 
the controllability matrix of the {\it original} plant system 
\eqref{linear dynamics} and \eqref{measurement}. 
Second, $\tilde{v}\in{\rm Ker}({\cal O}_y)$ is expressed by 
\[
    [M_1 C, M_1C_K]
     \left[ \begin{array}{cc} 
       A & \Sigma C^\top \Sigma C_K \\
       B_K M_1 C & A_K + B_K M_1 C_K \\
      \end{array}\right]^k
         \left[ \begin{array}{c} 
             v \\
             0 \\
         \end{array}\right]=0
\]
for all $k\geq 0$. 
This is equivalent to $M_1CA^kv = 0$, $\forall k\geq 0$, meaning 
that $v\in{\rm Ker}({\cal O}_y)$ for the original plant system.

Now we prove the theorem. 
Suppose that the original plant system \eqref{linear dynamics} and 
\eqref{measurement} does not have a QND variable; 
hence for any variable $\hat r=v^\top\hat x$, the vector $v$ satisfies 
$v\in{\rm Ker}({\cal C}_{\hat{\cal W}}^\top)^c$ or 
$v\in{\rm Range}({\cal O}_y^\top)^c$ for the original plant system. 
In particular, since the unobservability property does not depend on the 
choice of a specific coordinate, the latter condition is equivalently converted 
to $v\in{\rm Ker}({\cal O}_y)$. 
But as proven above, these two conditions are equivalent to 
$\tilde{v}\in{\rm Ker}({\cal C}_{\hat{\cal W}}^\top)^c$ or 
$\tilde{v}\in{\rm Ker}({\cal O}_y)$ for the closed-loop system; 
that is, the closed-loop system does not have a QND variable of the 
form $\hat r=v^\top\hat x=\tilde{v}^\top\hat x_e$. 
Thus the following result is obtained.

{\bf Theorem 2:} 
If the original plant system does not have a QND variable, then, 
any type-1 MF control cannot generate a QND variable in the 
closed-loop system.

%%%%%%%%%%%%%%%%%%%%%%%%%%%%%%%%%%%%%%%%%%%

\subsection{DFS}

Finally we prove the no-go theorem for generating a DFS via the type-1 
MF control. 
Let us assume that the closed-loop dynamics 
\eqref{type 1 closed-loop dynamics QND} with the output field 
\[
      \hat{\cal W}^{\rm out}
        =[C, C_K]
          \left[ \begin{array}{c} 
             \hat x \\
             x_K \\
          \end{array}\right]
      + \hat{\cal W}
\]
contains a DFS composed of ``purely quantum" variables of the form 
$\hat r=v^\top \hat x=\tilde{v}^\top\hat x_e$. 
Then, it follows from the statement below Eq.~\eqref{def of DFS alg} that 
$\tilde{v}\in{\rm Ker}({\cal C}_{\hat{\cal W}}^\top)$ and 
$\tilde{v}\in{\rm Ker}({\cal O}_{\hat{\cal W}^{\rm out}})$ hold. 
As proven in the QND case, the first condition equivalently leads to 
$v\in{\rm Ker}({\cal C}_{\hat{\cal W}}^\top)$ for the original plant 
system \eqref{linear dynamics} and \eqref{linear output}. 
Also in almost the same way we can prove that the second condition 
is equivalent to $v\in{\rm Ker}({\cal O}_{\hat{\cal W}^{\rm out}})$ 
for the original plant system. 
These two conditions on $v$ mean that the original plant system 
\eqref{linear dynamics} and \eqref{linear output} has a DFS, thus 
the contraposition yields the following theorem.

{\bf Theorem 3:} 
If the original plant system does not have a DFS, then, any type-1 
MF control cannot generate a DFS in the closed-loop system.

%%%%%%%%%%%%%%%%%%%%%%%%%%%%%%%%%%%%%%%%%%%%%
%%%%%%%%%%%%%%%%%%%%%%%%%%%%%%%%%%%%%%%%%%%%%
%%%%%%%%%%%%%%%%%%%%%%%%%%%%%%%%%%%%%%%%%%%%%

\section{The no-go theorems: type-2 case}

\begin{figure}[t]
\centering 
\includegraphics[width=6.9cm]{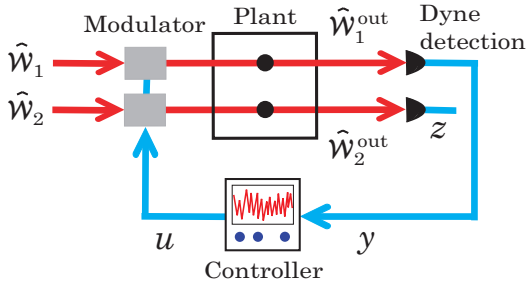}
\caption{
General configuration of the type-2 MF control. 
}
\label{MIMO general FB}
\end{figure}

In the type-1 case, it is assumed that at most all the plant's output fields 
can be used for feedback control and they are equally evaluated. 
For example, in the type-1 BAE case, the BA noise $\hat{\cal P}$ must not 
appear in {\it all} the elements of $y$. 
But it is sometimes more reasonable to give different roles to the 
output fields; 
such a control schematic in the MF case is illustrated in 
Fig.~\ref{MIMO general FB}, which we call the {\it type-2 control} 
configuration. 
In this case, at most all the components of $\hat{\cal W}_1^{\rm out}$ 
can be used for feedback control, while those of $\hat{\cal W}_2^{\rm out}$ 
are for evaluation; 
that is, they will be measured to extract some information about the system 
or will be kept untouched for later use. 
For instance, we attempt to design a MF control based on the measurement 
of $\hat{\cal W}_1^{\rm out}$, so that the BA noise does not appear in 
the measurement output of $\hat{\cal W}_2^{\rm out}$. 
However, we will see that such a MF control strategy does not work to 
achieve any of the control goals. 
That is, in this section, the type-2 no-go theorems in Table~I will be proven.

%%%%%%%%%%%%%%%%%%%%%%%%%%%%%%%%%%%%%%%%%%%

\subsection{The closed-loop system with type-2 MF}

As in the case of type-1 control, we study the situation where the 
feedback control is performed by modulating the input fields. 
The plant system driven by the modulated fields obeys the following 
dynamical equation:
\begin{eqnarray*}
& & \hspace*{0em}
    \frac{d\hat x}{dt} = A \hat x 
       + \Sigma C_1^\top \Sigma (\hat {\cal W}_1 + u_1)
       + \Sigma C_2^\top \Sigma (\hat {\cal W}_2 + u_2), 
\nonumber \\ & & \hspace*{0em}
    \hat {\cal W}_1^{\rm out} = C_1\hat x + \hat {\cal W}_1 + u_1,~~
    \hat {\cal W}_2^{\rm out} = C_2\hat x + \hat {\cal W}_2 + u_2. 
\end{eqnarray*}
$u_1$ and $u_2$ are the vectors of control signals that represent the 
time-varying amplitude of the input fields $\hat {\cal W}_1$ and 
$\hat {\cal W}_2$, respectively. 
Note that in general the size of $C_1$ and $C_2$ need not to be equal. 
The output field $\hat {\cal W}_1^{\rm out}$ is measured by a set of 
dyne detectors, which yield 
\[
    y = M\hat {\cal W}_1^{\rm out} 
       = MC_1\hat x + M\hat {\cal W}_1 + Mu_1. 
\]
$M$ is the symplectic matrix, representing which quadratures 
of $\hat {\cal W}_1^{\rm out}$ is measured. 
The measurement result $y(t)$ is sent to a classical feedback controller 
of the form 
\[
    \frac{dx_K}{dt}=A_K x_K + B_K y,~~
    u_1 = C_{K1} x_K,~~
    u_2 = C_{K2} x_K.
\]
Note that $u_2$ is allowed to contain the direct term from $y$, i.e. 
$u_2 = C_{K2} x_K+D_K y$; 
but this modification does not change the results shown below, thus for 
simplicity we assume $D_K=0$. 
Combining all the above equations, we end up with the closed-loop dynamics 
of $\hat x_e = [\hat x^\top, x_K^\top]^\top$: 
\begin{eqnarray}
\label{type 2 closed-loop dynamics QND}
& & \hspace*{-2em}
    \frac{d\hat x_e}{dt}
     =\left[ \begin{array}{cc} 
       A & \Sigma C_1^\top \Sigma C_{K1} + \Sigma C_2^\top \Sigma C_{K2}  \\
       B_K M C_1  &  A_K + B_K M C_{K1}\\
      \end{array}\right] \hat x_e
\nonumber \\ & & \hspace*{2em}
     \mbox{}
     + \left[ \begin{array}{c} 
                 \Sigma C_1^\top \Sigma \\
                 B_K M \\
               \end{array}\right]\hat{\cal W}_1
     + \left[ \begin{array}{c} 
                 \Sigma C_2^\top \Sigma \\
                 0 \\
               \end{array}\right]\hat{\cal W}_2. 
\end{eqnarray}
There are two kinds of output signals of the system. 
One is $y(t)$, which is used for feedback control. 
Due to the direct control term, it is now of the form 
\begin{equation}
\label{type 2 closed-loop output 1}
    y = [MC_1,~MC_{K1}] \hat x_e
             + M\hat{\cal W}_1.
\end{equation}
The other one is used for evaluation, which is obtained by measuring 
the second output field $\hat{\cal W}_2^{\rm out}$: 
\begin{equation}
\label{type 2 closed-loop output 2}
    z = M_1\hat{\cal W}_2^{\rm out}
      = [M_1C_2,~M_1C_{K2}] \hat x_e
      + \hat{\cal Q},
\end{equation}
where we have defined $\hat{\cal Q}=M_1\hat{\cal W}_2$.

%%%%%%%%%%%%%%%%%%%%%%%%%%%%%%%%%%%%%%%%%%%

\subsection{BAE}

The goal of BAE is to evade the BA noise so that it does not appear in 
the output signal \eqref{type 2 closed-loop output 2}. 
Now $\hat{\cal Q}=M_1\hat{\cal W}_2$ is the unavoidable shot noise and 
$\hat{\cal P}:=M_2 \hat{\cal W}_2$ is the BA noise, where the matrices 
satisfy Eq.~\eqref{M1 M2 conditions}. 
Note that the noise term of the closed-loop system 
\eqref{type 2 closed-loop dynamics QND} can be expressed by 
\begin{eqnarray*}
& & \hspace*{-1em}
     \mbox{noise term of Eq.~\eqref{type 2 closed-loop dynamics QND}}
\nonumber \\ & & \hspace*{-0.9em} 
     = \left[ \begin{array}{c} 
                 \Sigma C_1^\top \Sigma \\
                 B_K M \\
               \end{array}\right]\hat{\cal W}_1
     + \left[ \begin{array}{c} 
                 \Sigma C_2^\top \Sigma M_1^\top\\
                 0 \\
               \end{array}\right]\hat{\cal Q}
     + \left[ \begin{array}{c} 
                 \Sigma C_2^\top \Sigma M_2^\top\\
                 0 \\
               \end{array}\right]\hat{\cal P}. 
\nonumber
\end{eqnarray*}
Also the original system without control is given by 
\begin{eqnarray}
\label{original type 2 BAE}
& & \hspace*{-1em}
        \frac{d\hat x}{dt} = A \hat x 
           + \Sigma C_1^\top \Sigma \hat {\cal W}_1
           + \Sigma C_2^\top\Sigma(M_1^\top\hat{\cal Q}+M_2^\top\hat{\cal P}),
\nonumber \\ & & \hspace*{-0.8em}
        y = MC_1\hat x + M\hat {\cal W}_1,~~ 
        z = M_1C_2\hat x+\hat {\cal Q}.
\end{eqnarray}

We start with the assumption that BAE holds for the closed-loop 
system \eqref{type 2 closed-loop dynamics QND} and 
\eqref{type 2 closed-loop output 2}. 
In terms of the transfer function, this means that 
$\Xi_{\hat{\cal W}_1\rightarrow z}^{(fb)}[s]=0$ and 
$\Xi_{\hat{\cal P}\rightarrow z}^{(fb)}[s]=0$ are satisfied for all $s$, 
for this system (see Eq.~\eqref{def of BAE}). 
Thus, the Laplace transform of $z(t)$ is given by 
\begin{eqnarray}
& & \hspace*{-1em}
   z[s] = \Xi_{\hat{\cal Q}\rightarrow z}^{(fb)}[s]\hat{\cal Q}[s]
        + \Xi_{\hat{\cal P}\rightarrow z}^{(fb)}[s]\hat{\cal P}[s]
        + \Xi_{\hat{\cal W}_1\rightarrow z}^{(fb)}[s]\hat{\cal W}_1[s]
\nonumber \\ & & \hspace*{0.8em}
        = \Xi_{\hat{\cal Q}\rightarrow z}^{(fb)}[s]\hat{\cal Q}[s]. 
\nonumber
\end{eqnarray}
Let us now focus on the Laplace transform of $y(t)$: 
\[
   y[s] = \Xi_{\hat{\cal Q}\rightarrow y}^{(fb)}[s]\hat{\cal Q}[s]
        + \Xi_{\hat{\cal P}\rightarrow y}^{(fb)}[s]\hat{\cal P}[s]
        + \Xi_{\hat{\cal W}_1\rightarrow y}^{(fb)}[s]\hat{\cal W}_1[s]. 
\]
Both $z[s]$ and $y[s]$ are vectors of classical numbers, hence all their 
components commute with each other; 
i.e., $zy^\top-(yz^\top)^\top=0$ holds. 
Then, since in the Laplace domain the CCRs are represented by 
$[\hat{\cal Q}_j, \hat{\cal P}_k]=\delta_{jk}i/2s$, 
$[\hat{\cal Q}_j, \hat{\cal Q}_k]=0$, and 
$[\hat{\cal Q}_j, \hat{\cal W}_{1,k}]=0$, we have 
\begin{eqnarray}
& & \hspace*{-2em}
   zy^\top-(yz^\top)^\top
\nonumber \\ & & \hspace*{-1em}
    = \Xi_{\hat{\cal Q}\rightarrow z}^{(fb)}\hat{\cal Q}\hat{\cal P}^\top
        \big(\Xi_{\hat{\cal P}\rightarrow y}^{(fb)}\big)^\top
     - \Big[ \Xi_{\hat{\cal P}\rightarrow y}^{(fb)}\hat{\cal P}\hat{\cal Q}^\top
        \big(\Xi_{\hat{\cal Q}\rightarrow z}^{(fb)}\big)^\top \Big]^\top
\nonumber \\ & & \hspace*{-1em}
   = \Xi_{\hat{\cal Q}\rightarrow z}^{(fb)}
      \Big[\hat{\cal Q}\hat{\cal P}^\top 
      - \big(\hat{\cal P}\hat{\cal Q}^\top\big)^\top \Big]
        \big(\Xi_{\hat{\cal P}\rightarrow y}^{(fb)}\big)^\top
\nonumber \\ & & \hspace*{-1em}
    = \frac{i}{2s}\Xi_{\hat{\cal Q}\rightarrow z}^{(fb)}
                \big(\Xi_{\hat{\cal P}\rightarrow y}^{(fb)}\big)^\top
    =0. 
\nonumber 
\end{eqnarray}
But Eq.~\eqref{type 2 closed-loop output 2} clearly indicates that 
$\Xi_{\hat{\cal Q}\rightarrow z}^{(fb)}[s]$ is invertible for all $s$, hence we 
conclude $\Xi_{\hat{\cal P}\rightarrow y}^{(fb)}[s]=0,~\forall s$. 
This equivalently leads to the following set of equalities: 
\begin{eqnarray}
& & \hspace*{-1.7em}
     [MC_1, MC_{K1}]
     \left[ \begin{array}{cc} 
       A & \Sigma C_1^\top \Sigma C_{K1} + \Sigma C_2^\top \Sigma C_{K2}  \\
       B_K M C_1  &  A_K + B_K M C_{K1}\\
      \end{array}\right]^k
\nonumber \\ & & \hspace*{3em}
     \times
     \left[ \begin{array}{c} 
                 \Sigma C_2^\top \Sigma M_2^\top \\
                 0 \\
               \end{array}\right]
     =0,~~~\forall k\geq 0. 
\nonumber 
\end{eqnarray}
Likewise the proof in the type-1 case, we have 
\begin{equation}
\label{type 2 BAE proof}
   MC_1 A^k \Sigma C_2^\top \Sigma M_2^\top=0,~~~\forall k\geq 0,
\end{equation}
which implies that the original system \eqref{original type 2 BAE} satisfies 
$\Xi_{\hat{\cal P}\rightarrow y}^{(o)}[s]=0$, $\forall s$. 
Now the BAE condition 
$\Xi_{\hat{\cal P}\rightarrow z}^{(fb)}[s]=0$, $\forall s$ is expressed 
in the state space representation by 
\begin{eqnarray}
& & \hspace*{-1.3em}
    [M_1C_2, M_1C_{K2}]
      \left[ \begin{array}{cc} 
       A & \Sigma C_1^\top \Sigma C_{K1} + \Sigma C_2^\top \Sigma C_{K2}  \\
       B_K M C_1  &  A_K + B_K M C_{K1}\\
      \end{array}\right]^k 
\nonumber \\ & & \hspace*{6em}
      \times
      \left[ \begin{array}{c} 
            \Sigma C_2^\top \Sigma M_2^\top \\
            0 \\
        \end{array}\right]
       =0,~~\forall k\geq 0. 
\nonumber 
\end{eqnarray}
Then, using Eq.~\eqref{type 2 BAE proof}, we deduce 
\[
     M_1C_2 A^k \Sigma C_2^\top \Sigma M_2^\top = 0,~~\forall k\geq 0. 
\]
Hence, for the original system \eqref{original type 2 BAE}, the 
transfer function from $\hat {\cal P}$ to $z$ is zero; 
i.e., $\Xi_{\hat{\cal P}\rightarrow z}^{(o)}[s]=0,~\forall s$.

We finally prove $\Xi_{\hat{\cal W}_1\rightarrow z}^{(o)}[s]=0,~\forall s$. 
The above result implies 
$z^{(o)}[s]
=\Xi_{\hat{\cal W}_1\rightarrow z}^{(o)}[s]\hat {\cal W}_1[s]
+\Xi_{\hat{\cal Q}\rightarrow z}^{(o)}[s]\hat {\cal Q}[s]$. 
Moreover, from Eq.~\eqref{type 2 BAE proof} we have 
$\Xi_{\hat{\cal P}\rightarrow y}^{(o)}[s]=0,~\forall s$, which leads to 
$y^{(o)}[s]
=\Xi_{\hat{\cal W}_1\rightarrow y}^{(o)}[s]\hat {\cal W}_1[s]
+\Xi_{\hat{\cal Q}\rightarrow y}^{(o)}[s]\hat {\cal Q}[s]$. 
Then, since both $y^{(o)}[s]$ and $z^{(o)}[s]$ are c-numbers, 
we have 
\begin{eqnarray}
& & \hspace*{-2em}
   y^{(o)}z^{(o)}\mbox{}^\top-(z^{(o)}y^{(o)}\mbox{}^\top)^\top
    = \Xi_{\hat{\cal W}_1\rightarrow y}^{(o)}\hat{\cal W}_1\hat{\cal W}_1^\top
        \big(\Xi_{\hat{\cal W}_1\rightarrow z}^{(o)}\big)^\top
\nonumber \\ & & \hspace*{6em}
     - \Big[ \Xi_{\hat{\cal W}_1\rightarrow z}^{(o)}\hat{\cal W}_1\hat{\cal W}_1^\top
        \big(\Xi_{\hat{\cal W}_1\rightarrow y}^{(o)}\big)^\top \Big]^\top
\nonumber \\ & & \hspace*{1em}
   = \Xi_{\hat{\cal W}_1\rightarrow y}^{(o)}
      \Big[\hat{\cal W}_1\hat{\cal W}_1^\top 
      - \big(\hat{\cal W}_1\hat{\cal W}_1^\top\big)^\top \Big]
        \big(\Xi_{\hat{\cal W}_1\rightarrow z}^{(o)}\big)^\top
\nonumber \\ & & \hspace*{1em}
   = \frac{i}{2s}\Xi_{\hat{\cal W}_1\rightarrow y}^{(o)} \Sigma 
                \big(\Xi_{\hat{\cal W}_1\rightarrow z}^{(o)}\big)^\top=0. 
\nonumber 
\end{eqnarray}
Note now that $\Xi_{\hat{\cal W}_1\rightarrow z}^{(o)}[s]$ does not 
depend on the matrix $M$, representing which quadratures of 
$\hat{\cal W}_1^{\rm out}$ are measured. 
This means that the above equality holds for other choice of 
measurement, say $\tilde{y}=\tilde{M}\hat{\cal W}_1$. 
Thus we have 
\begin{eqnarray}
& & \hspace*{-1em}
    \left[ \begin{array}{c} 
            \Xi_{\hat{\cal W}_1\rightarrow y}^{(o)} \\
            \Xi_{\hat{\cal W}_1\rightarrow \tilde{y}}^{(o)} \\
        \end{array}\right]
         \Sigma \big(\Xi_{\hat{\cal W}_1\rightarrow z}^{(o)}\big)^\top
    =\left[ \begin{array}{c} 
            M \\
            \tilde{M} \\
        \end{array}\right]
         \Xi_{\hat{\cal W}_1\rightarrow \hat{\cal W}_1^{\rm out}}^{(o)}
           \Sigma \big(\Xi_{\hat{\cal W}_1\rightarrow z}^{(o)}\big)^\top
\nonumber \\ & & \hspace*{9.7em}
    =0.
\nonumber 
\end{eqnarray}
$\tilde{M}$ is chosen so that $[M^\top, \tilde{M}^\top]$ is invertible. 
Because $\Xi_{\hat{\cal W}_1\rightarrow \hat{\cal W}_1^{\rm out}}^{(o)}[s]$ 
is also invertible, 
$\Xi_{\hat{\cal W}_1\rightarrow z}^{(o)}[s]=0$, $\forall s$. 
Together with the above result $\Xi_{\hat{\cal P}\rightarrow z}^{(o)}[s]=0$, 
$\forall s$, this means that BAE holds for the original plant system 
\eqref{original type 2 BAE}. 
Consequently, we have the following result:

{\bf Theorem 4:} 
If the original plant system does not have the BAE property, then, 
any type-2 MF control cannot realize BAE for the closed-loop system.

%%%%%%%%%%%%%%%%%%%%%%%%%%%%%%%%%%%%%%%%%%%

\subsection{QND}

The idea for the proof is the same as that taken in the type-1 case. 
Again, a QND variable is of the form 
$\hat r=v^\top \hat x=\tilde{v}^\top\hat x_e$ with 
$\tilde{v}=[v^\top, 0^\top]^\top$. 
Now the closed-loop system is given by Eqs. 
\eqref{type 2 closed-loop dynamics QND}, \eqref{type 2 closed-loop output 1}, 
and \eqref{type 2 closed-loop output 2}, showing that it is subjected to the 
input noise field $[\hat{\cal W}_1^\top, \hat{\cal W}_2^\top]^\top$ and 
it generates the measurement outputs $[y^\top, z^\top]^\top$. 
Thus by definition $\hat r$ is a QND variable iff 
$\tilde{v}\in{\rm Ker}({\cal C}_{\hat{\cal W}_1}^\top)\cap
{\rm Ker}({\cal C}_{\hat{\cal W}_2}^\top)$ and 
$\tilde{v}\in{\rm Range}({\cal O}_y^\top)\cup{\rm Range}({\cal O}_z^\top)$. 
The former condition means that 
\begin{eqnarray}
& & \hspace*{-2em}
    [v^\top,~0^\top]
    \left[ \begin{array}{cc} 
      A & \Sigma C_1^\top \Sigma C_{K1} + \Sigma C_2^\top \Sigma C_{K2}  \\
       B_K M C_1  &  A_K + B_K M C_{K1}\\
    \end{array}\right]^k
\nonumber \\ & & \hspace*{2em}
    \times
    \left[ \begin{array}{cc} 
                 \Sigma C_1^\top \Sigma & \Sigma C_2^\top \Sigma \\
                 B_K M & 0\\
               \end{array}\right]=0,~~\forall k\geq 0. 
\nonumber
\end{eqnarray}
This is equivalent to 
$v^\top A^k\Sigma C_1^\top \Sigma=0$ and 
$v^\top A^k\Sigma C_2^\top \Sigma=0$ for all $k\geq 0$; 
that is, $v\in{\rm Ker}({\cal C}_{\hat{\cal W}_1}^\top)\cap
{\rm Ker}({\cal C}_{\hat{\cal W}_2}^\top)$ holds for the original plant 
system \eqref{original type 2 BAE}. 
(Note $\hat{\cal W}_2=M_1^\top\hat{\cal Q}+M_2^\top\hat{\cal P}$.) 
Related to the latter one, let us consider the condition 
$\tilde{v}\in{\rm Ker}({\cal O}_y)\cap{\rm Ker}({\cal O}_z)$. 
This is expressed by 
\begin{eqnarray}
& & \hspace*{-1em}
    \left[ \begin{array}{cc} 
                 M C_1 & M C_{K1} \\
                 M_1 C_2 & M_1C_{K2} \\
               \end{array}\right]
\nonumber \\ & & \hspace*{0em}
    \times 
    \left[ \begin{array}{cc} 
      A & \Sigma C_1^\top \Sigma C_{K1} + \Sigma C_2^\top \Sigma C_{K2}  \\
       B_K M C_1  &  A_K + B_K M C_{K1}\\
    \end{array}\right]^k
        \left[ \begin{array}{c} 
                 v \\
                 0 \\
               \end{array}\right]=0,
\nonumber
\end{eqnarray}
for all $k\geq 0$, which equivalently leads to 
\[
     \left[ \begin{array}{cc} 
                 M C_1 \\
                 M_1 C_2 \\
               \end{array}\right] A^k v = 0,~~\forall k\geq 0.
\]
Thus $v\in{\rm Ker}({\cal O}_y)\cap{\rm Ker}({\cal O}_z)$ holds 
for the original plant system \eqref{original type 2 BAE}. 
From the same discussion as that in Sec.~IV-C together with the 
above results, we obtain the following no-go theorem:

{\bf Theorem 5:} 
If the original plant system does not have a QND variable, then, 
any type-2 MF control cannot generate a QND variable in the 
closed-loop system.

%%%%%%%%%%%%%%%%%%%%%%%%%%%%%%%%%%%%%%%%%

\subsection{DFS}

Let us assume that the closed-loop system 
\eqref{type 2 closed-loop dynamics QND} with the output fields 
$\hat{\cal W}_1^{\rm out}$ and $\hat{\cal W}_2^{\rm out}$, 
which now satisfy 
\[
     \left[ \begin{array}{c} 
          \hat{\cal W}_1^{\rm out} \\
          \hat{\cal W}_2^{\rm out} \\
        \end{array}\right]
      = \left[ \begin{array}{cc} 
            C_1 & C_{K1} \\
            C_2 & C_{K2} \\
         \end{array}\right]
         \left[ \begin{array}{c} 
          \hat x \\
          x_K \\
        \end{array}\right]
      + \left[ \begin{array}{c} 
           \hat{\cal W}_1 \\
           \hat{\cal W}_2 \\
        \end{array}\right], 
\]
contains a DFS. 
Equivalently, it contains a subsystem that is uncontrollable 
w.r.t. $\hat{\cal W}_1$ and $\hat{\cal W}_2$ and 
unobservable w.r.t. $\hat{\cal W}_1^{\rm out}$ and 
$\hat{\cal W}_2^{\rm out}$. 
As before, a variable contained in the DFS is of the form 
$\hat r= v^\top \hat x=\tilde{v}^\top\hat x_e$. 
Then, first, the uncontrollability condition leads to the same results 
as in the QND case, i.e. $v\in{\rm Ker}({\cal C}_{\hat{\cal W}_1}^\top)\cap
{\rm Ker}({\cal C}_{\hat{\cal W}_2}^\top)$ holds for the original plant 
system \eqref{original type 2 BAE}. 
Further, it is immediate to see that the unobservability condition yields 
$C_1A^k v=0$ and $C_2A^k v=0$ for all $k\geq 0$. 
Consequently, $v\in{\rm Ker}({\cal C}_{\hat{\cal W}_1}^\top)
\cap{\rm Ker}({\cal C}_{\hat{\cal W}_2}^\top)$ and 
$v\in{\rm Ker}({\cal O}_{\hat{\cal W}_1^{\rm out}})
\cap{\rm Ker}({\cal O}_{\hat{\cal W}_2^{\rm out}})$ hold for the original 
plant system. 
Thus we have the following result.

{\bf Theorem 6:} 
If the original plant system does not have a DFS, then, any type-2 
MF control cannot generate a DFS in the closed-loop system.

%%%%%%%%%%%%%%%%%%%%%%%%%%%%%%%%%%%%%%%%%%%%%
%%%%%%%%%%%%%%%%%%%%%%%%%%%%%%%%%%%%%%%%%%%%%
%%%%%%%%%%%%%%%%%%%%%%%%%%%%%%%%%%%%%%%%%%%%%

\section{Coherent feedback realizations: type-1 case}

Here we turn our attention to the CF control and in what follows will see that, 
as shown in Table~I, it has a capability of achieving the control goals, 
BAE, QND, and DFS. 
That is, as mentioned in Sec.~I, these are situations where a quantum 
device has a clear advantage over a classical one. 
This section is devoted to prove the results in the type-1 CF case.

%%%%%%%%%%%%%%%%%%%%%%%%%%%%%%%%%%%%%%%%%%%%%

\subsection{The closed-loop system with type-1 CF}

\begin{figure}[t]
\centering 
\includegraphics[width=4.1cm]{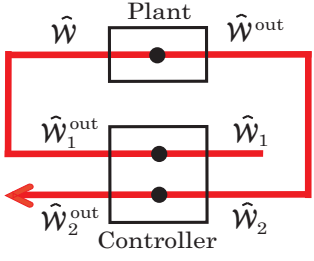}
\caption{
A general configuration of the type-1 CF control.}
\label{type 1 CF fig}
\end{figure}

The plant system is given by Eqs.~\eqref{linear dynamics} and 
\eqref{linear output} with input $\hat{\cal W}$ and output 
$\hat{\cal W}^{\rm out}$. 
In the type-1 control configuration, as described in Sec.~IV, at most 
all the components of $\hat{\cal W}^{\rm out}$ can be used for feedback, 
and also at most all the components of $\hat{\cal W}$ can be controlled. 
A CF controller is constructed by directly connecting another fully quantum 
system to the plant system by a feedback way. 
This means that, in the type-1 CF case, $\hat{\cal W}^{\rm out}$ is 
connected to the controller's input and the controller's output is connected 
to $\hat{\cal W}$, without involving any measurement process. 
The CF control configuration satisfying this setting, which avoids 
self-interaction of the fields, is depicted in Fig.~\ref{type 1 CF fig}. 
The controller has two kinds of input-output fields, and its system 
equation is given by 
\begin{eqnarray}
\label{type 1 CF}
& & \hspace*{-1em}
   \frac{d\hat x_K}{dt} = A_K \hat x_K 
       + \Sigma C_1^\top \Sigma \hat {\cal W}_1
       + \Sigma C_2^\top \Sigma \hat {\cal W}_2, 
\nonumber \\ & & \hspace*{-1em}
    \hat {\cal W}_1^{\rm out} = C_1\hat x_K + \hat {\cal W}_1,~~
    \hat {\cal W}_2^{\rm out} = C_2\hat x_K + \hat {\cal W}_2, 
\end{eqnarray}
where $A_K=\Sigma(G_K + C_1^\top\Sigma C_1/2 + C_2^\top\Sigma C_2/2)$. 
The CF control is constructed by 
\begin{equation}
\label{type 1 CF connection}
     \hat {\cal W}_2=\hat {\cal W}^{\rm out},~~~
     \hat {\cal W}=\hat {\cal W}_1^{\rm out}.
\end{equation}
This condition imposes the size of $C_1$ and $C_2$ to be equal, although 
they are not necessarily of full rank. 
Note that more generally a scattering process from e.g. 
$\hat {\cal W}^{\rm out}$ to $\hat {\cal W}_2$ can be introduced, but 
here it is not necessary. 
Combining Eqs.~\eqref{linear dynamics}, \eqref{linear output}, 
\eqref{type 1 CF}, and \eqref{type 1 CF connection}, we obtain the 
dynamical equation of the closed-loop system: 
\begin{equation}
\label{closed loop system type 1 CF}
\hspace*{-0.5em}
    \frac{d\hat x_e }{dt}
       = A_e \hat x_e
       + \Sigma C_e^\top \Sigma \hat {\cal W}_1,~~
    \hat {\cal W}_2^{\rm out} 
      = C_e \hat x_e + \hat {\cal W}_1,
\end{equation}
where $\hat x_e=[\hat x^\top, \hat x_K^\top]^\top$, 
$A_e=\Sigma(G_e + C_e^\top \Sigma C_e/2)$, $C_e=[C, C_1+C_2]$, and 
\[
    G_e 
      = \left[ \begin{array}{cc} 
          G & C^\top \Sigma C_1/2 - C^\top \Sigma C_2/2 \\
          \star 
               & ~G_K + C_1^\top\Sigma^\top C_2/2 + C_2^\top\Sigma C_1/2 \\
        \end{array}\right]. 
\]
$\star$ denotes the symmetric elements of $G_e$.

%%%%%%%%%%%%%%%%%%%%%%%%%%%%%%%%%%%%%%%%%%%%%

\subsection{BAE}

\begin{figure}[t]
\centering 
\includegraphics[width=8.5cm]{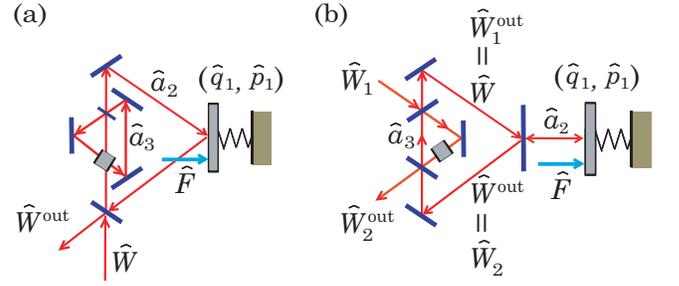}
\caption{
(a) Direct interaction scheme achieving BAE for the opto-mechanical oscillator, 
proposed by Tsang and Caves \cite{Tsang2010}.
(b) Equivalent realization via the type-1 CF control. 
}
\label{Tsang scheme}
\end{figure}

Let us assume that we can engineer a CF controller satisfying 
$C_1+C_2=0$. 
Then the closed-loop system \eqref{closed loop system type 1 CF} takes the 
following form:
\begin{eqnarray}
\label{type 1 CF direct int}
& & \hspace*{0em}
    \frac{d \hat x_e}{dt}
        =  \left[ \begin{array}{cc} 
               A & \Sigma C^\top \Sigma C_1 \\
               \Sigma C_1^\top \Sigma^\top C & \Sigma G_K \\
             \end{array}\right]
             \hat x_e
        + \left[ \begin{array}{c} 
                \Sigma C^\top \Sigma  \\
                0 \\
            \end{array}\right] \hat {\cal W}_1,
\nonumber \\ & & \hspace*{0em}
    \hat {\cal W}_2^{\rm out} 
      = [C,~0] \hat x_e
       + \hat {\cal W}_1.
\end{eqnarray}
The structure of this equation shows that, notably, the controller is directly 
coupled to the plant, yet there is no direct interaction between the field 
and the controller. 
This system configuration is called the {\it direct interaction}, meaning that 
an additional quantum device is prepared and is directly coupled to the 
plant system, not through input/output fields; 
hence the system \eqref{type 1 CF direct int} is a CF-based realization 
of the direct interaction.

Here we study the opto-mechanical oscillator described in Sec.~II-C (ii), 
as a plant system. 
Since this system has one input-output field, the control configuration 
must be of type-1. 
Also it is easy to verify that this system does not satisfy BAE, and further, 
it does not have a QND variable. 
The goal is to design a CF controller such that BAE is realized for the 
closed-loop system toward high-precision detection of the unknown force 
$\hat F$. 
For this purpose, we take the CF scheme described above, leading to 
Eq.~\eqref{type 1 CF direct int}. 
The controller is single mode with variable $\hat x_K=[\hat q_3, \hat p_3]^\top$, 
and it has two input fields $\hat{W}_1=[\hat Q_1, \hat P_1]^\top$ 
and $\hat{W}_2=[\hat Q_2, \hat P_2]^\top$. 
The controller's system matrices are chosen so that they satisfy 
\[
     \Sigma C_1^\top \Sigma^\top C
      = \left[ \begin{array}{cc|cc} 
                  & 0 &    & 0 \\
                0&    & g &    \\
            \end{array}\right],~~
      \Sigma G_K = 
         \left[ \begin{array}{cc} 
                  & -\omega \\
                \omega &    \\
            \end{array}\right],
\]
which leads to 
\begin{equation}
\label{CF type 1 BAE controller matrices}
     C_1 = - C_2 
      = \frac{g}{\sqrt{2\gamma}}
          \left[ \begin{array}{cc} 
                0 & 0 \\
                1 & 0 \\
            \end{array}\right],~~
     G_K = 
         \left[ \begin{array}{cc} 
                -\omega & 0 \\
                0 & -\omega \\
            \end{array}\right].
\end{equation}
Physical implementation of the controller specified by these matrices 
will be discussed in the end of this subsection. 
Together with the term $\hat F$, which directly acts on $\hat p_1$, 
the dynamics of the closed-loop system is given by 
\begin{equation}
\label{type 1 CF opto-mechanics}
\hspace*{0.0em}
    \frac{d\hat x_e }{dt}
       = A_e \hat x_e + B_e \hat {W}_1 + b_f \hat F,~~
    \hat {W}_2^{\rm out} 
       = C_e \hat x_e + \hat {W}_1,
\end{equation}
where 
\begin{eqnarray*}
& & \hspace*{-1em}
    A_e 
    =\left[ \begin{array}{cc|cc|cc} 
        & 1/m & & 0 & &  \\
       -m\omega^2 & & \kappa & & & \\ \hline
        & 0 & -\gamma &  & & 0 \\
       \kappa & &  & -\gamma & g & \\ \hline
        & & & 0 & & -\omega \\
        & & g & & \omega &  \\
               \end{array}\right],~
     B_e = C_e^\top,
\nonumber \\ & & \hspace*{-1em}
     C_e
        = \sqrt{2\gamma}
          \left[ \begin{array}{cc|cc|cc} 
             0 &    & 1 &    & 0 &   \\
             & 0 &    & 1 &    & 0 \\
          \end{array}\right],~~
      b_f=[0, 1, 0, 0, 0, 0]^\top.
\end{eqnarray*}
Since $\hat Q_2^{\rm out}$ does not contain any information about 
$\hat F$, we need to measure $\hat P_2^{\rm out}$, implying that 
the output signal is given by $y=M\hat {W}_2^{\rm out}
=\hat P_2^{\rm out}$ with $M=[0,~1]$, i.e. 
\begin{equation}
\label{type 1 CF opto-mechanics output}
     y = c_y \hat x_e + \hat P_1
        = \sqrt{2\gamma}[0, 0, 0, 1, 0, 0]\hat x_e + \hat P_1.
\end{equation}
The set of equations \eqref{type 1 CF opto-mechanics} and 
\eqref{type 1 CF opto-mechanics output} is exactly the same as that of 
the modified opto-mechanical oscillator proposed by Tsang and Caves 
\cite{Tsang2010}, which is shown in Fig.~\ref{Tsang scheme}~(a). 
Notably, this system realizes BAE measurement for detecting $\hat F$; 
in fact, with the choice $g=\kappa/\sqrt{m\omega}$ the transfer function 
from the BA noise $\hat Q_1$ to the output $y=\hat P_2^{\rm out}$ 
takes zero: 
\[
      y[s] = \frac{\sqrt{2\gamma}\kappa/m}
                        {(s+\gamma)(s^2+\omega^2)}\hat P_1[s]
               + \frac{s-\gamma}{s+\gamma} \hat F[s]. 
\]
Thus by injecting a $\hat P_1$-squeezed light field (i.e. by reducing the 
noise of $\hat P_1$), in principle we can detect $\hat F$ with better 
accuracy compared to the case without BAE. 
A detailed investigation of this BAE scheme in a practical setting was 
recently reported in \cite{Heurs}.

Recall now that the system \eqref{type 1 CF opto-mechanics} and 
\eqref{type 1 CF opto-mechanics output} is constructed by a CF control. 
That is, in a constructive way, we have proven that the type-1 
CF control can realize BAE.

Lastly, let us consider an optical implementation of the above CF controller. 
The form of $C_1$ (or $C_2$) in Eq.~\eqref{CF type 1 BAE controller 
matrices} represents the so-called QND interaction of the controller and 
the field $\hat{W}_1$ (or $\hat{W}_2$), which can be physically implemented 
though in a nontrivial way \cite{WisemanPRA93}. 
The controller's Hamiltonian specified by $G_K$ in Eq.~\eqref{CF type 1 
BAE controller matrices} simply expresses the optical phase shift. 
Consequently, a detuned optical cavity coupled to two input-output fields 
via QND interactions, illustrated in Fig.~\ref{Tsang scheme} (b), is one 
possible physical realization of the CF controller proposed here. 
Note that its practical implementation is harder than that of the system 
given in \cite{Tsang2010}. 
But apart from such difficulty, again, what should be emphasized here 
is the fact that the type-1 CF control is capable of realizing BAE.

%%%%%%%%%%%%%%%%%%%%%%%%%%%%%%%%%%%%%%%%%%%%%

\subsection{QND}

Let us continue to examine the above CF-controlled opto-mechanical 
oscillator \eqref{type 1 CF opto-mechanics} and 
\eqref{type 1 CF opto-mechanics output}; 
actually we here show that this system contains QND variables, by 
proving Eq.~\eqref{def of QND alg}, which is now 
${\rm Ker}({\cal C}^\top_{\hat{W}_1}) \cap 
{\rm Range}({\cal O}_y^\top) \neq \emptyset$.

First, if $g=\kappa/\sqrt{m\omega}$, the range of the controllability matrix 
${\cal C}_{\hat{W}_1} = [B_e, A_eB_e, A_e^2B_e]$ is spanned by the 
following independent vectors:
\[
          \left[ \begin{array}{c} 
                   0 \\
                   0 \\
                   1 \\
                   0 \\
                   0 \\
                   0 \\
               \end{array}\right],~~
            \left[ \begin{array}{c} 
                   0 \\
                   0 \\
                   0 \\
                   1 \\
                   0 \\
                   0 \\
               \end{array}\right],~~
             \left[ \begin{array}{c} 
                   0\\
                   \kappa\\
                   0\\
                   0\\
                   0\\
                   g\\
               \end{array}\right],~~
             \left[ \begin{array}{c} 
                   \kappa/m \\
                   0 \\
                   0 \\
                   0 \\
                   -g\omega \\
                   0\\
               \end{array}\right].
\]
Note that $A_e^kB_e~(k\geq 3)$ does not anymore produce an independent 
vector. 
Clearly, 
\[
     v_1 = [0, -g, 0, 0, 0, \kappa]^\top,~~~
     v_2 = [g\omega, 0, 0, 0, \kappa/m,0]^\top
\]
are contained in ${\rm Ker}({\cal C}_{\hat{W}_1}^\top)$. 
Next, the kernel of the observability matrix 
${\cal O}_y=[c_y^\top, A_e^\top c_y^\top, \ldots]^\top$ is spanned by 
\[
        \left[ \begin{array}{c} 
                   0 \\
                   0 \\
                   1 \\
                   0 \\
                   0 \\
                   0 \\
               \end{array}\right],~~
             \left[ \begin{array}{c} 
                   0\\
                   g\omega \\
                   0\\
                   0\\
                   0\\
                   \kappa/m \\
               \end{array}\right],~~
             \left[ \begin{array}{c} 
                   -g \\
                   0 \\
                   0 \\
                   0 \\
                   \kappa \\
                   0\\
               \end{array}\right].
\]
But they are orthogonal to both $v_1$ and $v_2$, meaning that 
$v_1$ and $v_2$ are contained in ${\rm Range}({\cal O}_y^\top)$. 
Consequently, we find that 
$v_1, v_2 \in {\rm Ker}({\cal C}_{\hat{W}_1}^\top) \cap 
{\rm Range}({\cal O}_y^\top)$. 
Thus 
\[
     \hat q' = v_1^\top \hat x_e
                  = -g\hat p_1 + \kappa \hat p_3,~~
     \hat p' = v_2^\top \hat x_e
                  = g\omega\hat q_1 + \frac{\kappa}{m} \hat q_3
\]
are uncontrollable w.r.t.~$\hat{W}_1$ and observable w.r.t. $y$ 
(see the discussion around Eq.~\eqref{def of QND alg}); 
that is, $\hat q'$ and $\hat p'$ are QND variables generated by the CF control. 
Indeed, they are subjected to the dynamical equation of the form 
\begin{equation}
\label{classical subsystem}
     \frac{d\hat q'}{dt}
       = \frac{\omega}{m} (g^2 m^2 \omega^2 + \kappa^2) \hat p',~~
     \frac{d\hat p'}{dt}
       = -\frac{\omega}{m}
                   (g^2 + \kappa^2) \hat q' + \hat F,
\end{equation}
which clearly shows that $(\hat q', \hat p')$ are free from $\hat{W}_1$.

Here an interesting by-product is obtained. 
It is easy to see $[\hat q'(t), \hat p'(t)]=0$, $\forall t$. 
Together with the fact that $(\hat q', \hat p')$ are independent from 
other variables, this means that they are essentially classical variables 
which are detectable from the output field. 
In general, if a quantum system contains a subsystem whose variables 
are all commutative, then it is called a {\it classical subsystem} 
\cite{Tsang2012}; 
thus we now found that the CF-controlled opto-mechanical system 
\eqref{type 1 CF opto-mechanics} contains a classical subsystem 
\eqref{classical subsystem}.

%%%%%%%%%%%%%%%%%%%%%%%%%%%%%%%%%%%%%%%%%%%%%

\subsection{DFS}

To show that the type-1 CF control has capability of generating a DFS, 
let us return to the general closed-loop system 
\eqref{closed loop system type 1 CF}. 
Suppose now that the original plant system \eqref{linear dynamics} 
and \eqref{linear output} does not have a DFS, and further that 
a quantum controller with parameters $C_1=C_2=C/2$ and $G_K=G$ 
can be engineered. 
Hence, the plant and the controller have the same number of modes. 
Then Eq.~\eqref{closed loop system type 1 CF} takes the following form: 
\begin{eqnarray}
\label{type 1 CF DFS}
& & \hspace*{-0.5em}
       \frac{d \hat x_e}{dt}
          = A_e\hat x_e + B_e \hat {\cal W}_1,~~
       \hat {\cal W}_2^{\rm out} = C_e\hat x_e + \hat {\cal W}_1,
\nonumber \\ & & \hspace*{-0.5em}
       A_e  =  \left[ \begin{array}{cc} 
               A & \Sigma C^\top \Sigma C/2 \\
               \Sigma C^\top \Sigma C/2 & A \\
             \end{array}\right],~
       B_e = \left[ \begin{array}{c} 
                    \Sigma C^\top \Sigma \\
                    \Sigma C^\top \Sigma \\
                  \end{array}\right],
\nonumber \\ & & \hspace*{-0.5em}
        C_e = [C,~C].
\end{eqnarray}

Now we prove that this system contains a DFS, i.e. a subsystem that is 
uncontrollable w.r.t. $\hat {\cal W}_1$ and unobservable w.r.t. 
$\hat {\cal W}_2^{\rm out}$. 
First, for the vector $v_e=[v^\top, -v^\top]^\top$ with $v$ an arbitrary 
$2n$-dimensional real vector, we have
\[
     v_e^\top A_e^k B_e 
      = [v^\top(\Sigma G)^k,~ -v^\top(\Sigma G)^k]
           \left[ \begin{array}{c} 
                    \Sigma C^\top \Sigma \\
                    \Sigma C^\top \Sigma \\
           \end{array}\right] = 0, 
\]
for all $k \geq 0$. 
Hence, $v_e \in {\rm Ker}({\cal C}_{\hat{\cal W}_1}^\top)$ holds 
with ${\cal C}_{\hat{\cal W}_1}=[B_e, A_eB_e, A_e^2B_e, \ldots]$ 
the controllability matrix. 
Also, 
\[
     C_e A_e^k v_e 
      = [C,~C]
           \left[ \begin{array}{c} 
                    (\Sigma G)^k v \\
                    -(\Sigma G)^k v \\
           \end{array}\right] = 0, ~~\forall k\geq 0
\]
holds, implying $v_e \in {\rm Ker}({\cal O}_{\hat{\cal W}_2^{\rm out}})$ 
with ${\cal O}_{\hat{\cal W}_2^{\rm out}}$ the observability matrix 
${\cal O}_{\hat{\cal W}_2^{\rm out}}
=[C_e^\top, A_e^\top C_e^\top, \ldots]^\top$. 
Consequently, $v_e$ satisfies 
$v_e \in {\rm Ker}({\cal C}_{\hat{\cal W}_1}^\top) \cap 
{\rm Ker}({\cal O}_{\hat{\cal W}_2^{\rm out}})$. 
This means, as discussed above Eq.~\eqref{def of DFS alg 2}, that 
$v_e^\top\hat x_e = v^\top \hat x - v^\top \hat x_K$ is uncontrollable 
and unobservable, hence this is the variable of a DFS generated by the CF 
control. 
Note that $2n$ independent vectors $(v_1, \ldots v_{2n})$ can be taken to 
construct $v_e^{(i)}=[v_i^\top, -v_i^\top]^\top$. 
Thus this DFS is composed of $2n$ variables 
$\{v_i^\top \hat x - v_i^\top \hat x_K\}_{i=1,\ldots,2n}$.

%%%%%%%%%%%%%%%%%%%%%%%%%%%%%%%%%%%%%%%%%%%%%
%%%%%%%%%%%%%%%%%%%%%%%%%%%%%%%%%%%%%%%%%%%%%
%%%%%%%%%%%%%%%%%%%%%%%%%%%%%%%%%%%%%%%%%%%%%

\section{Coherent feedback realizations: type-2 case}

In this section we study the type-2 CF control for realizing BAE, QND, 
and DFS. 
As in the type-1 case, a specific system achieving each control goal will 
be shown.

%%%%%%%%%%%%%%%%%%%%%%%%%%%%%%%%%%%%%%%%%%%%

\subsection{The closed-loop system with type-2 CF}

\begin{figure}[t]
\centering 
\includegraphics[width=5.2cm]{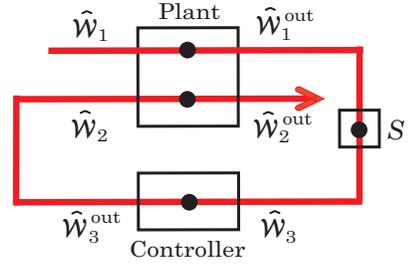}
\caption{
A general configuration of the type-2 CF control.}
\label{type 2 CF}
\end{figure}

As explained in Sec.~V, the type-2 control means that two roles are 
given to the output fields of the plant system; 
one is for feedback control, and the other one is for evaluation. 
Hence the system to be controlled is 
\begin{eqnarray}
\label{type 2 general plant}
& & \hspace*{0em}
    \frac{d\hat x}{dt} = A \hat x 
       + \Sigma C_1^\top \Sigma \hat {\cal W}_1
       + \Sigma C_2^\top \Sigma \hat {\cal W}_2, 
\nonumber \\ & & \hspace*{0em}
    \hat {\cal W}_1^{\rm out} = C_1\hat x + \hat {\cal W}_1,~~
    \hat {\cal W}_2^{\rm out} = C_2\hat x + \hat {\cal W}_2. 
\end{eqnarray}
For designing a CF controller, there are some variation in its structure. 
Here we particularly consider the CF control configuration illustrated 
in Fig.~\ref{type 2 CF}; 
that is, the controller has a single kind of input-output fields that are 
directly connected to the plant's input and output fields. 
For a general type-2 CF control configuration, see \cite{JamesTAC2008}. 
Note also that, in our case, $C_1$ and $C_2$ are of the same size, 
although they are not necessarily of full rank. 
Hence the dynamics of the CF controller is given by 
\begin{equation*}
    \frac{d\hat x_K}{dt} = A_K\hat x_K 
       + \Sigma C_K^\top \Sigma \hat {\cal W}_3,~~~
    \hat {\cal W}_3^{\rm out} = C_K \hat x_K + \hat {\cal W}_3,
\end{equation*}
where $A_K=\Sigma(G_K+C_K^\top\Sigma C_K/2)$, and the feedback 
connection is realized by 
\[
     \hat {\cal W}_3=S\hat {\cal W}_1^{\rm out},~~~
     \hat {\cal W}_2=\hat {\cal W}_3^{\rm out}.
\]
Here $S$ is an orthogonal and symplectic matrix representing a scatting 
process from $\hat {\cal W}_1^{\rm out}$ to $\hat {\cal W}_3$. 
Combining the above equations yields the dynamical equation of the 
closed-loop system; 
\begin{equation}
\label{type 2 CF whole network}
\hspace{0.5em}
    \frac{d\hat x_e}{dt} 
       = A_e \hat x_e
       + \Sigma C_e^\top \Sigma S \hat {\cal W}_1,~~
    \hat {\cal W}_2^{\rm out} 
      = C_e \hat x_e + S \hat {\cal W}_1,
\end{equation}
where $\hat x_e = [\hat x^\top, \hat x_K^\top]^\top$ and 
\begin{eqnarray*}
& & \hspace*{-1em}
     A_e=\Sigma [G_e+C_e^\top\Sigma C_e/2],~~~
     C_e=[SC_1+C_2,~C_K], 
\nonumber \\ & & \hspace*{-1em}
     G_e = \left[ \begin{array}{cc} 
                    G + (C_2^\top \Sigma S C_1 + C_1^\top S^\top \Sigma^\top C_2)/2 & 
                          \star \\
                    C_K^\top \Sigma(SC_1-C_2)/2 & G_K \\
                  \end{array}\right]. 
\nonumber
\end{eqnarray*}
$\star$ denotes the symmetric elements of $G_e$.

%%%%%%%%%%%%%%%%%%%%%%%%%%%%%%%%%%%%%%%%%%%%%

\subsection{BAE}

\begin{figure}[t]
\centering 
\includegraphics[width=5.6cm]{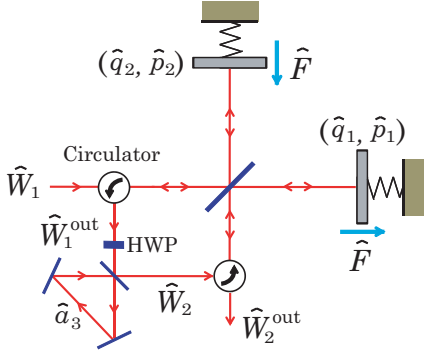}
\caption{
Michelson's interferometer with type-2 CF. 
The CF controller is an optical cavity with coupling constant $\epsilon$ 
and detuning $\alpha$. HWP: half wave plate. 
}
\label{GW with CF}
\end{figure}

To demonstrate that the type-2 CF is capable of realizing BAE, we here 
study the Michelson's interferometer as a plant system, which is described 
in Sec. II-C (iii) with Fig.~\ref{examples} (c). 
The system is composed of two oscillators driven by an unknown force 
$\hat F$ along opposite directions. 
The oscillators' dynamical motion is described by 
Eq.~\eqref{GW without control}, which is specified by the following 
system matrices: 
$G={\rm diag}\{m\omega^2, 1/m,m\omega^2, 1/m\}$ and 
\[
       C_1 = \sqrt{\lambda}
                  \left[ \begin{array}{cc|cc} 
                      & 0 &  & 0 \\
                     1 &  & 1 & 
                   \end{array}\right],~
       C_2 = \sqrt{\lambda}
                  \left[ \begin{array}{cc|cc} 
                      & 0 &  & 0 \\
                     1 &  & -1 & 
                   \end{array}\right]. 
\]
This system works as a sensor for detecting the force $\hat F$; 
but as explained before, the noise power of the output signal is bounded 
from below by the SQL \eqref{SQL}. 
Hence the purpose here is to design a CF controller that realizes 
BAE and as a result beats the SQL. 
Actually, the plant system has two input-output ports, hence it 
can be treated within the type-2 CF control framework.

Here we consider the CF configuration described in the previous subsection. 
That is, $\hat{W}_1^{\rm out}$ and $\hat{W}_2$ are optically connected 
through CF. 
In particular, as a CF controller, let us take a single input-output 
optical cavity, whose dynamical equation is specified by the following 
matrices: 
\[
       G_K = \left[ \begin{array}{cc} 
                      \alpha & 0 \\
                      0 & \beta
                   \end{array}\right],~~
        C_K = \sqrt{2\epsilon}
                   \left[ \begin{array}{cc} 
                      1 & 0 \\
                      0 & 1
                   \end{array}\right],~~
        S = \left[ \begin{array}{cc} 
                      0 & 1 \\
                      -1 & 0
                   \end{array}\right], 
\]
where $\epsilon$ is the coupling constant between the field and the cavity 
mode. 
Later we will set $\alpha=\beta$, which thus represents the detuning. 
$S$ represents a phase shift acting on the input optical field in the form 
$\hat A_3 = -i \hat A_1^{\rm out}$. 
Thus the closed-loop system is a 3-modes single input-output linear system, 
depicted in Fig.~\ref{GW with CF}.

With the above setup, the closed-loop system \eqref{type 2 CF whole network} 
takes the following form: 
\begin{eqnarray}
\label{system matrices type 2}
& & \hspace*{-1.1em}
       \frac{d \hat x_e}{dt}
          = A_e\hat x_e + B_e \hat W_1' + b_f\hat F,~~
       \hat W_2^{\rm out} = C_e\hat x_e + \hat W_1',
\nonumber \\ & & \hspace*{-1.1em}
    A_e
     =\left[ \begin{array}{cc|cc|cc} 
       & 1/m & & 0 & & 0 \\
      \lambda-m\omega^2 & & \lambda & & \sqrt{2\lambda\epsilon} & \\ \hline
       & 0 & & 1/m & & 0 \\
      -\lambda & & -\lambda-m\omega^2 & & -\sqrt{2\lambda\epsilon} & \\ \hline
      -\sqrt{2\lambda\epsilon} & & -\sqrt{2\lambda\epsilon} & & -\epsilon & \beta \\
       & 0 & & 0 & -\alpha & -\epsilon \\
               \end{array}\right],
\nonumber \\ & & \hspace*{-1.1em}
    B_e = \Sigma C_e^\top \Sigma = [b_1,~b_2]
     = \left[ \begin{array}{cc} 
                 0 & 0\\
                 \sqrt{\lambda} & -\sqrt{\lambda} \\ \hline
                 0 & 0 \\
                 -\sqrt{\lambda} & -\sqrt{\lambda} \\ \hline
                 -\sqrt{2\epsilon} & 0 \\
                 0 & -\sqrt{2\epsilon} \\
               \end{array}\right],
\nonumber \\ & & \hspace*{-1.1em}
     b_f = 
     \left[ \begin{array}{cc|cc|cc} 
           0 & 1 & 0 & -1 & 0 & 0 \\
     \end{array}\right]^\top, 
\nonumber \\ & & \hspace*{-1.1em}
     C_e
      = \left[ \begin{array}{c} 
                  c_1^\top \\
                  c_2^\top 
          \end{array}\right]
      = \left[ \begin{array}{cc|cc|cc} 
           \sqrt{\lambda} & 0 & \sqrt{\lambda} & 0 & \sqrt{2\epsilon} & 0 \\
           \sqrt{\lambda} & 0 & -\sqrt{\lambda} & 0 & 0 & \sqrt{2\epsilon} \\
         \end{array}\right],
\nonumber \\ & & \hspace*{-1.1em}
      \hat{W}_1' = S\hat{W}_1=[\hat P_1, -\hat Q_1]^\top.
\end{eqnarray}

Let us seek the parameters $(\alpha, \beta, \epsilon)$ that achieve BAE. 
First, it is easy to see $c_1^\top A_e^k b_f=0,~\forall k\geq 0$, 
or equivalently ${\rm Ker}({\cal C}_{\hat{F}}^\top)^c\cap
{\rm Range}({\cal O}_{\hat Q_2^{\rm out}}^\top) = \emptyset$; 
that is, $\hat Q_2^{\rm out}$ does not contain any information about 
$\hat F$. 
Thus we measure 
\begin{equation}
\label{CF type 2 BAE y}
     y=\hat P_2^{\rm out}=c_2^\top\hat x_e - \hat Q_1,
\end{equation}
implying that $\hat Q_1$ is the shot noise while $\hat P_1$ is the BA noise. 
Thus the parameters should be chosen so that the BAE condition 
\eqref{def of BAE alg} i.e. 
${\rm Ker}({\cal C}_{\hat{P}_1}^\top)^c\cap{\rm Range}({\cal O}_y^\top) 
= \emptyset$ is satisfied, which is carried out by examining the 
equivalent condition \eqref{def of BAE alg 2}: 
$c_2^\top A_e^k b_1=0,~\forall k\geq 0$. 
The case $k=0$ is already satisfied. 
To see the case $k\geq 1$, let us focus on 
\[
   A_e b_1 = \left[ \begin{array}{c} 
             \sqrt{\lambda}/m \\
             0 \\ \hline
             -\sqrt{\lambda}/m \\
             0 \\ \hline
             -\epsilon\sqrt{2\epsilon} \\
             \alpha\sqrt{2\epsilon} \\
          \end{array}\right],~~
   A_e^2 b_1 = \left[ \begin{array}{c} 
               0 \\
               -(\omega^2+2\epsilon^2)\sqrt{\lambda} \\ \hline
               0 \\
               (\omega^2+\epsilon^2/2)\sqrt{\lambda} \\ \hline
               (\alpha \beta+\epsilon^2)\sqrt{2\epsilon} \\
               0 \\
             \end{array}\right],
\]
where the proportional part to $b_1$ are subtracted. 
Then, the condition is satisfied if we impose $c_2^\top A_e b_1=0$ 
and $A_e^2 b_1\propto b_1$, which yield
\[
    \frac{\lambda}{m}+\alpha\epsilon=0,~~~
    \omega^2 + 2\epsilon^2 = \alpha\beta + \epsilon^2. 
\]
Let us especially take the parameter $\alpha=\beta<0$, implying that 
the CF controller is an optical cavity with negative detuning $\alpha$. 
The parameters are then explicitly given by 
\begin{equation}
\label{BAE condition type 2}
    \epsilon 
      = \frac{\sqrt{2}\lambda}
            {m\sqrt{\omega^2 + \sqrt{\omega^4+4\lambda^2/m^2}}},~~
    \alpha = \frac{-\lambda}{m\epsilon}. 
\end{equation}
When $\omega \ll 1$, they are approximated by $\sqrt{\lambda/m}$ and 
$-\sqrt{\lambda/m}$, respectively. 
Actually under the condition \eqref{BAE condition type 2}, the output is 
described in the Laplace domain by 
\begin{eqnarray*}
& & \hspace*{-1em}
    y[s] = -\frac{s^3 - \epsilon s^2 + \omega^2 s 
                        + \epsilon(\omega^2 + 2\epsilon^2) }
                             {(s^2+\omega^2)(s+\epsilon)}\hat{Q}_1[s] 
\nonumber \\ & & \hspace*{4.5em}
   \mbox{}
             + \frac{2\sqrt{\lambda}}{m(s^2+\omega^2)} \hat{F}[s],
\end{eqnarray*}
which is free from the BA noise $\hat{P}_1[s]$. 
As expected, this BAE measurement beats the SQL and enables 
high-precision detection of $\hat F$. 
To see this fact, let us evaluate the power spectrum density of the noise. 
As seen before, $\hat F$ induces the oscillators's position shift $\hat g$ 
in the Fourier domain $s=i\Omega$ by 
$\hat F[i\Omega]=-mL\Omega^2 \hat g[i\Omega]$. 
Then, under the assumption $\omega \ll \Omega$, the normalized signal 
is given by 
\[
   \tilde{y}[i\Omega]
    =\frac{y[i\Omega]}{2\sqrt{\lambda}L}
    = \hat{g}[i\Omega] 
           - \frac{i\Omega^3 - \epsilon\Omega^2 - 2\epsilon^3}
                  {2\sqrt{\lambda}L\Omega^2 (i\Omega + \epsilon) }
                       \hat{Q}_1[i\Omega]. 
\]
Using $\epsilon=\sqrt{\lambda/m}$ we obtain
\[
   S[i\Omega]=\mean{|\tilde{y}-\hat{g}|^2}
      = \Big( \frac{\lambda}{m^2L^2\Omega^4}
                + \frac{1}{4\lambda L^2} \Big) \mean{|\hat{Q}_1|^2}, 
\]
which has the same form as that of the non-controlled scheme in 
Eq.~\eqref{SQL}, except that the BA noise is replaced by the shot noise. 
Therefore, by injecting a $\hat Q_1$-squeezed light field into the 
{\it first} input port (i.e. the bright port), we can realize a broadband 
noise reduction below the SQL \eqref{SQL} in the output noise power. 
It should be noted again that, without squeezing of the input field, the 
output noise power of the CF-controlled interferometer having 
the BAE property reproduces the SQL. 
This means that achieving BAE itself does not necessarily result in the 
increased force sensitivity; 
in fact we need to combine the BAE property and squeezing of the input.

Note that, while we have found a CF controller achieving BAE for 
high-precision detection of $\hat F$ below the SQL, the result obtained 
here does not mean to emphasize that the proposed schematic is an 
alternative configuration for gravitational wave detection. 
Actually, the schematic is very different from several effective methods, 
particularly in that the second output port is not anymore a dark port. 
Hence the amplitude component must be subtracted from the output field, 
which though cannot be carried out perfectly; 
thus the above-described ideal detection of $\hat g$ below the SQL would 
be a difficult task in a practical situation. 
Rather the main purpose here is to prove the capability of a type-2 CF 
controller for realizing BAE. 
Also, as demonstrated above, it is remarkable that the problem for designing 
BAE can be solved, by a system theoretic approach based on the 
controllability/observability notion; 
this approach might shed a new light on the engineering problems for 
gravitational wave detection.

%%%%%%%%%%%%%%%%%%%%%%%%%%%%%%%%%%%%%%%%%%%%%

\subsection{QND}

We here see that the closed-loop system studied in the previous subsection 
contains QND variables. 
Note that the original interferometer does not have a QND variable.

First let us calculate the controllability matrix 
${\cal C}_{\hat{W}'_1}=[B_e, A_eB_e, A_e^2 B_e, \ldots]$ with $A_e$ and 
$B_e$ given in Eq.~\eqref{system matrices type 2}. 
It was already seen that $b_1$ generates two dimensional subspace 
spanned by $b_1$ and $A_e b_1$, under the condition 
\eqref{BAE condition type 2}. 
Now, by further imposing $\alpha=\beta$, we have 
$A_e^2 b_2 = -\omega^2 b_2$, implying that 
${\rm Range}({\cal C}_{\hat{W}'_1})$ is spanned by 
\[
             \left[ \begin{array}{c} 
                   0 \\
                   \sqrt{\lambda} \\
                   0 \\
                   -\sqrt{\lambda} \\
                   -\sqrt{2\epsilon} \\
                   0 \\
               \end{array}\right],~ 
               \left[ \begin{array}{c} 
                   0 \\
                   \sqrt{\lambda} \\
                   0 \\
                   \sqrt{\lambda} \\
                   0 \\
                   \sqrt{2\epsilon} \\
               \end{array}\right],~ 
               \left[ \begin{array}{c} 
                   \sqrt{\lambda}/m \\
                   0 \\
                   -\sqrt{\lambda}/m \\
                   0 \\
                   -\epsilon\sqrt{2\epsilon} \\
                   \alpha\sqrt{2\epsilon} \\
               \end{array}\right],~ 
               \left[ \begin{array}{c} 
                   \sqrt{\lambda}/m \\
                   0 \\
                   \sqrt{\lambda}/m \\
                   0 \\
                   \beta\sqrt{2\epsilon} \\
                   -\epsilon\sqrt{2\epsilon} \\
               \end{array}\right]. 
\]
Hence ${\rm dim}{\rm Range}({\cal C}_{\hat{W}'_1})=4$. 
Let us take two independent vectors $v_1$ and $v_2$ spanning 
${\rm Ker}({\cal C}_{\hat{W}'_1}^\top)$; 
then $v_1^\top\hat x_e$ and $v_2^\top\hat x_e$ are not affected by 
the input field $\hat{W}'_1$. 
Moreover, these variables appear in the output signal \eqref{CF type 2 BAE y} 
as shown below. 
Actually we can prove that $c_2$ and $A_e^\top c_2$ are both independent 
to the above four vectors, implying
\[
    {\rm Range}({\cal C}_{\hat{W}'_1})\oplus 
      {\rm span}\{c_2,~A_e^\top c_2 \} = {\mathbb R}^6.
\]
Thus ${\rm Range}({\cal C}_{\hat{W}'_1})\cup
{\rm Range}({\cal O}_y^\top) = {\mathbb R}^6$ holds, which further leads to 
${\rm Range}({\cal C}_{\hat{W}'_1})^c \subseteq{\rm Range}({\cal O}_y^\top)$. 
Consequently, we find $v_1, v_2\in{\rm Range}({\cal O}_y^\top)$, meaning 
that $v_1^\top\hat x_e$ and $v_2^\top\hat x_e$ appear in $y$ and thus 
they are QND variables. 
That is, the type-2 CF controller described in Sec.~VII-B has capability 
of generating QND variables.

%%%%%%%%%%%%%%%%%%%%%%%%%%%%%%%%%%%%%%%%%%%%%

\subsection{DFS}

Lastly we again study a general CF-controlled system 
\eqref{type 2 CF whole network}; 
suppose that the plant system \eqref{type 2 general plant} satisfies 
$C_1=C_2=C/2$ and does not contain a DFS. 
Further, let us choose a type-2 CF controller with system matrices 
$G_K=G$ and $C_K=C$, which is directly connected to the plant (i.e. $S=I$). 
Then Eq.~\eqref{type 2 CF whole network} takes exactly 
the same form as Eq.~\eqref{type 1 CF DFS}, which contains a DFS. 
Therefore, this type-2 CF controller has ability to generate a DFS.

%{\it Remark:} 
%The CF-controlled interferometer depicted in Fig.~\ref{GW system DFS} 
%has a DFS, hence this is another demonstration of a CF control generating 
%a DFS. 

%\begin{figure}[t]
%\centering 
%\includegraphics[width=5.5cm]{GW_DFS_CF.eps}
%\caption{
%Another configuration of the type-2 CF control for Michelson's interferometer. 
%This system contains a DFS. 
%}
%\label{GW system DFS}
%\end{figure} 

%%%%%%%%%%%%%%%%%%%%%%%%%%%%%%%%%%%%%%%%%%%%%
%%%%%%%%%%%%%%%%%%%%%%%%%%%%%%%%%%%%%%%%%%%%%
%%%%%%%%%%%%%%%%%%%%%%%%%%%%%%%%%%%%%%%%%%%%%

\section{Conclusion and future works}

This paper has given some general answers to the question about whether 
or not measurement should be involved in the feedback structure for 
controlling a quantum system. 
That is, for a general linear quantum system, we have obtained 
the no-go theorems stating that the control goal, realization of BAE, 
QND, or DFS, cannot be achieved by any MF control; 
on the other hand, for each control goal, we have found an example of 
CF control accomplishing the task. 
From the viewpoint that MF is essentially a classical operation on the 
system while CF is a fully quantum one, these results imply that 
BAE, QND, and DFS are genuine quantum objectives that cannot be 
realized by any feedback-based classical operation.

The key idea to obtain all the results is the following system theoretic 
characterizations of BAE, QND, and DFS, which are also summarized in 
Fig.~\ref{Goals}: 
\begin{eqnarray*}
%\label{def conclusion algebraic}
& & \hspace*{-1em}
       \mbox{{\bf BAE:}}~~~
       {\rm Ker}({\cal C}_{\hat{\cal P}}^\top)^c\cap{\rm Range}({\cal O}_y^\top) 
       = \emptyset, 
\nonumber \\ & & \hspace*{-1em}
       \mbox{{\bf QND:}}~~~
       {\rm Ker}({\cal C}_{\hat{\cal W}}^\top) \cap 
    {\rm Range}({\cal O}_y^\top) \neq \emptyset, 
\nonumber \\ & & \hspace*{-1em}
       \mbox{{\bf DFS:}}~~~
       {\rm Ker}({\cal C}_{\hat{\cal W}}^\top) \cap 
       {\rm Range}({\cal O}_{\hat{\cal W}^{\rm out}}^\top)^c \neq \emptyset.
\end{eqnarray*}
Now we should remember the following equivalent characterizations 
in terms of transfer functions: 
\begin{eqnarray*}
%\label{def conclusion transfer}
& & \hspace*{-1em}
       \mbox{{\bf BAE:}}~~~ 
          \Xi_{\hat P\rightarrow y}[s]=0,~\forall s,
\nonumber \\ & & \hspace*{-1em}
       \mbox{{\bf QND:}}~~~ 
       \Xi_{\hat{\cal W} \rightarrow \hat x_2'}[s]=0,~\forall s~~\&~~
       \Xi_{\hat x_2' \rightarrow y}[s]\neq 0,~\exists s,
\nonumber \\ & & \hspace*{-1em}
       \mbox{{\bf DFS:}}~~~ 
       \Xi_{\hat{\cal W} \rightarrow \hat x_2'}[s]=0,~\forall s~~\&~~
       \Xi_{\hat x_2' \rightarrow \hat{\cal W}^{\rm out}}[s]=0,~\forall s.
\end{eqnarray*}
Although in this paper these characterizations are not fully used except 
Sec.~V-B, they will serve as powerful tools in quantum device engineering 
in a practical situation. 
In fact, in reality due to several experimental imperfections, it is often the 
case that the controllability/observability matrix becomes of full rank, 
and thus the perfect achievement of the above geometric conditions 
cannot be expected. 
Nonetheless, the functional approach based on the transfer function allows 
us to obtain an approximate solution of those problems. 
For instance for the BAE case, even if 
${\rm Ker}({\cal C}_{\hat{\cal P}}^\top)^c\cap{\rm Range}({\cal O}_y^\top) 
= \emptyset$ or equivalently $\Xi_{\hat P\rightarrow y}[s]=0,~\forall s$ 
is never satisfied, an approximate BAE measurement can be 
engineered by solving a minimization problem 
$\|\Xi_{\hat P\rightarrow y}[s]\|\rightarrow {\rm min}$. 
Actually, in the history of classical control, the so-called geometric 
control theory was first deeply investigated \cite{Wonham}, pursuing e.g. 
ideal disturbance decoupling. 
Later, towards wider applicability of the control theory, several functional 
approaches were developed \cite{Zhou Doyle book}; 
the linear quadratic Gaussian (LQG) control and $H^\infty$ control, 
which are respectively based on the minimization of the $H^2$ norm 
$\|\cdot\|_2$ and the $H^\infty$ norm $\|\cdot\|_\infty$ of a 
transfer function, are typical successful results. 
A notable fact is that, as mentioned in Sec.~I, recently quantum 
versions of those classical feedback control methods have been deeply 
developed. 
Therefore combination of the geometric and functional approaches 
will constitute a new methodology in the field of quantum control and 
information. 
Of course, under the evaluation of minimizing a norm of a transfer 
function, comparing MF and CF controls again becomes an open problem.

Another important direction of the future research is to extend the 
results to the nonlinear case. 
Actually the control goals, BAE, QND, DFS, are all essential as well 
in nonlinear systems, such as optical devices with high order nonlinearity, 
photonic crystal arrays, and coupled qubits networks. 
The strength of the input-output formalism \cite{GardinerBook,WallsMilburn} 
is in that it is applicable to a very wide class of such Markovian nonlinear 
systems. 
More precisely, for a general system that couples with $m$ probe/environment 
fields, its variable $\hat X(t)$ is governed by the following quantum 
stochastic differential equation: 
\begin{eqnarray}
& & \hspace*{-1em}
\label{general QSDE}
    \frac{d\hat X}{dt} 
      = i[\hat H, \hat X] 
         + \sum_{j=1}^m \big( 
           \hat L_j^* \hat X \hat L_j
            -\frac{1}{2} \hat L_j^* \hat L_j \hat X
              -\frac{1}{2} \hat X \hat L_j^* \hat L_j \big)
\nonumber \\ & & \hspace*{3em}
    \mbox{}
      + \sum_{j=1}^m \Big( 
          [\hat X, \hat L_j] \hat A_j^*
           -  [\hat X, \hat L_j^*] \hat A_j \Big), 
\end{eqnarray}
where $\hat H$ is the system Hamiltonian and $\hat L_j$ is the 
coupling operator. 
Also the $j$th output field satisfies 
\begin{equation}
\label{general output}
     \hat A^{\rm out}_j = \hat L_j + \hat A_j. 
\end{equation}
In fact, the nonlinear atomic ensemble dynamics \eqref{spin ensemble QSDE} 
is obtained by setting $\hat H=0$ and $\hat L=\sqrt{M}\hat J_z$ in 
Eq.~\eqref{general QSDE}. 
(Also, the linear system \eqref{linear dynamics} and \eqref{linear output} 
corresponds to the case $\hat H=\hat x^\top G\hat x/2$ and 
$\hat L_j=c_j ^\top\hat x$.) 
Very importantly, there exists a celebrated classical nonlinear systems 
and control theory \cite{Isidori,Nijmeijer}, that gives clear characterizations 
of controllability and observability notions even for nonlinear systems. 
Therefore it is expected that, by taking a similar approach shown in this paper, 
we can have a unified formalism of BAE, QND, and DFS for a general 
quantum nonlinear system \eqref{general QSDE} and \eqref{general output}. 
This should be very useful for systematic engineering of wider class of 
quantum information processing devices; 
but, as in the case discussed in the previous paragraph, comparison of 
MF and CF for nonlinear systems is also a nontrivial task. 
An interesting result along this direction was recently reported in 
\cite{Harris}; for the problem detecting a force driving a linear 
oscillator, a MF has clear advantage over the non-controlled system with 
an optimized estimator, only when the oscillator contains some nonlinearity.

%%%%%%%%%%%%%%%%%%%%%%%%%%%%%%%%%%%%%%%%%%%%%
%%%%%%%%%%%%%%%%%%%%%%%%%%%%%%%%%%%%%%%%%%%%%
%%%%%%%%%%%%%%%%%%%%%%%%%%%%%%%%%%%%%%%%%%%%%

\begin{acknowledgements}
This work was supported in part by JSPS Grant-in-Aid 
No. 40513289. 
The author acknowledges helpful discussions with I. R. Petersen. 
\end{acknowledgements}

%%%%%%%%%%%%%%%%%%%%%%%%%%%%%%%%%%%%%%%%%%%%%
%%%%%%%%%%%%%%%%%%%%%%%%%%%%%%%%%%%%%%%%%%%%%
%%%%%%%%%%%%%%%%%%%%%%%%%%%%%%%%%%%%%%%%%%%%%

\appendix

\section{Direct measurement feedback}

In this paper, from the standpoint comparing CF and MF, we assumed 
that a MF controller is given by a {\it dynamical} one with internal 
variable $x_K$ and that the control is carried out by modulating the 
plant's input fields. 
However, the control configuration is not limited to the dynamical one; 
the {\it direct (or proportional) measurement feedback} developed by 
Wiseman \cite{WisemanPRA94} is indeed the first proposal applying 
the classical feedback control in the quantum domain. 
As discussed in the literature (e.g. see \cite{WisemanBook}), an ideal 
MF control is actually effective in controlling the system; 
what is most notable here is the fact obtained in \cite{Wiseman1995}, 
clarifying that a direct MF can produce a QND variable, unlike the 
dynamical one. 
Let us here review this result.

The plant system is an optical cavity containing a $\chi^2$ nonlinear crystal, 
and further, the cavity mode can be directly controlled by a modulator. 
The output signal is obtained by measuring the amplitude quadrature of 
the output field. 
The system equations are then given by 
\[
    \frac{d\hat x}{dt}
       = \left[ \begin{array}{cc} 
                   -\kappa & 0 \\
                   0 & 0 \\
                  \end{array}\right]\hat x
         + \left[ \begin{array}{c} 
                 1 \\
                 0 \\
               \end{array}\right]u
         -\sqrt{\kappa} 
           \left[ \begin{array}{c} 
                   \hat Q \\
                   \hat P \\
                  \end{array}\right],~~
    y = \sqrt{\kappa} \hat q + \hat Q,
\]
where $\hat x = [\hat q, \hat p]^\top$ is the cavity mode quadratures, 
$u(t)$ is the control signal representing the amplitude modulation, 
and $\kappa$ is the coupling strength between the cavity and the probe 
field. 
Note that this modulation effect does not appear in the output. 
The direct feedback considered in \cite{Wiseman1995} is of the form 
$u=\sqrt{\kappa}y$, which enables us to modify the system dynamics 
so that $\hat x$ evolves in time with the following linear equation: 
\[
    \frac{d}{dt}
       \left[ \begin{array}{c} 
                 \hat q \\
                 \hat p \\
               \end{array}\right]
       = -\sqrt{\kappa} 
           \left[ \begin{array}{c} 
                   0 \\
                   1 \\
                  \end{array}\right] \hat P,~~~~
    y = \sqrt{\kappa} \hat q + \hat Q.
\]
Clearly, $\hat q$ is not disturbed by the noise while it appears in the output 
signal, implying that we can measure $\hat q$ without disturbing it. 
That is, $\hat q$ is a QND variable.

The above result means that the type-1 no-go theorem for QND does not 
hold, if an ideal direct MF can be employed. 
However, we should note a critical assumption that an ideal direct MF 
controller has infinite bandwidth. 
Hence let us further examine a practical case where the feedback circuit 
has a finite bandwidth and its dynamics is given by 
\begin{equation}
\label{appendix}
     \frac{dx_K}{dt}=-\frac{1}{\tau}x_K + \frac{1}{\tau}y,~~~
     u=\sqrt{\kappa}x_K, 
\end{equation}
where $\tau$ represents the time constant and $x_K$ is the internal variable 
of the circuit. 
Actually the transfer function from $y$ to $u$ is given by 
$\Xi_{y\rightarrow u}[s]=\sqrt{\kappa}/(1+\tau s)$, whose gain in the 
Fourier domain is computed as 
\[
     |\Xi_{y\rightarrow u}[i\Omega]|^2 
        = \frac{\kappa}{1+\tau^2 \Omega^2}. 
\]
The bandwidth is defined by $[-1/\tau, 1/\tau]$, in which more than 
half the power of the signal $y$ is allowed to pass through the circuit. 
This clearly shows that the MF is only available in the infinite 
bandwidth limit $\tau\rightarrow +0$. 
We can also see the finite bandwidth effect on the ideal QND variable 
$\hat q$ as follows; 
the combined system dynamics of the cavity and the circuit is given by 
\[
    \frac{d}{dt}
           \left[ \begin{array}{c} 
                 \hat q \\
                 x_K \\
               \end{array}\right]
       =  \left[ \begin{array}{cc} 
                   -\kappa & \sqrt{\kappa} \\
                   \sqrt{\kappa}/\tau & -1/\tau \\
           \end{array}\right]
           \left[ \begin{array}{c} 
                 \hat q \\
                 x_K \\
               \end{array}\right]
           + \left[ \begin{array}{c} 
                 -\sqrt{\kappa} \\
                 1/\tau \\
               \end{array}\right]\hat Q, 
\]
which yields 
\[
     \Xi_{\hat Q\rightarrow \hat q}[s] 
        = \frac{-\sqrt{\kappa} \tau}{(\kappa \tau+1)+\tau s}. 
\]
Thus, actually in the ideal limit $\tau\rightarrow +0$, the variable 
$\hat q$ becomes QND. 
In other words, a practical direct MF does not generate a QND variable. 
Note that controlling via the field modulation $\hat Q \rightarrow \hat Q + u$ 
together with the finite-bandwidth MF controller \eqref{appendix} is exactly 
the type-I MF, meaning that the no-go theorem is applied to this practical 
case. 
We should rather have an understanding that the controller \eqref{appendix} 
is an effective MF realizing an approximated QND variable in the 
scenario discussed in Sec.~VIII.

%%%%%%%%%%%%%%%%%%%%%%%%%%%%%%%%%%%%%%%%%%%%%
%%%%%%%%%%%%%%%%%%%%%%%%%%%%%%%%%%%%%%%%%%%%%
%%%%%%%%%%%%%%%%%%%%%%%%%%%%%%%%%%%%%%%%%%%%%

\end{document}